\documentclass[fleqn,usenatbib]{mnras}

\usepackage[T1]{fontenc} 
\usepackage{ulem}

\DeclareRobustCommand{\VAN}[3]{#2}
\let\VANthebibliography\thebibliography
\def\thebibliography{\DeclareRobustCommand{\VAN}[3]{##3}\VANthebibliography}

\usepackage{graphicx}	
\usepackage{amsmath}	
\usepackage{amssymb}	
\usepackage{movie15}
\usepackage{url}
\usepackage{newtxtext,newtxmath}
\let\oldAA\AA
\renewcommand{\AA}{\text{\normalfont\oldAA}}

\newcommand{\ms}{$M_{\star}$}

\newcommand{\mspipe}{$M_{\star}^{\rm P3D}$}
\newcommand{\msun}{M$_{\sun}$}
\newcommand{\re}{$R_{e}$}
\newcommand{\reg}{$R_{\rm e,g}$}
\newcommand{\rer}{$R_{\rm e,r}$}

\newcommand{\rmass}{$R_{\rm 50M}$}
\newcommand{\rlight}{$R_{\rm 50L}$}
\newcommand{\rlightr}{$R_{50r}$}
\newcommand{\tlb}{$t_{\rm lb}$}
\newcommand{\tq}{$t_{\rm quench}$}
\def\conc{\mbox{$C_{0.5-1.2}$}}

\title[Mass-to-light structural ratios of Ellipticals]{The differences between mass- and light-derived structural parameters over time for MaNGA Elliptical galaxies}
\author[H. Ibarra-Medel et al.]{
H. Ibarra-Medel,$^{1}$\thanks{E-mail: hjibarram@gmail.com (HIM)}
V. Avila-Reese,$^{2}$
I.~Lacerna,$^{3,4}$
A. Rodr\'iguez-Puebla,$^{2}$ \and
J.~A.~V\'azquez-Mata,$^{2,5}$
H.~M.~Hern\'andez-Toledo,$^{2}$
S.~F.~S\'anchez$^{2}$
\\
$^{1}$University of Illinois Urbana-Champaign, Department of Astronomy, 1002 W Green St, Urbana, Illinois, 61801, United States\\
$^{2}$Instituto de Astronom\'{\i}a, Universidad Nacional Aut\'onoma de M\'exico, A.P. 70-264, 04510 M\'exico D. F., M\'exico \\
$^{3}$Instituto de Astronom\'ia y Ciencias Planetarias, Universidad de Atacama, Copayapu 485, Copiap\'o, Chile\\
$^{4}$Millennium Institute of Astrophysics, Nuncio Monsenor Sotero Sanz 100, Of. 104, Providencia, Santiago, Chile \\
$^{5}$Departamento de F\'isica,  Facultad de Ciencias,  Universidad Nacional Aut\'onoma de M\'exico, A.P. 70-264, 04510 M\'exico D. F., M\'exico
}

\date{Accepted XXX. Received YYY; in original form ZZZ}

\pubyear{2015}

\begin{document}
\label{firstpage}
\pagerange{\pageref{firstpage}--\pageref{lastpage}}
\maketitle

\begin{abstract}

{We apply stellar population synthesis analysis to obtain spatially-resolved archaeological inferences for a large sample of ``red and dead'' Elliptical galaxies (Classical Ellipticals; CLEs) from the MaNGA/SDSS-IV DR15 survey.}
From their 2D stellar light and mass maps, we explore the differences between the radial mass and light distributions in the rest-frame bands $g,$ $r,$ and $i$ as functions of look-back time, \tlb, or redshift, $z$. We characterize these differences through
the ratios between the following mass- and light-derived global properties: sizes, concentrations, and effective surface densities. We find that the mass-to-light ratios of these properties change with \tlb, more the more massive the galaxies are. The CLE galaxy archaeological progenitors are, on average, less compact, concentrated, and dense in light than in mass as $z$ decreases. However, at later times, when also the {evolution of the progenitors becomes passive at all radii,}
there is an upturn in these trends and the differences between mass and light in compactness/concentration decrease towards $z\sim 0$. The trends in the ratios of mass to light sizes agree qualitatively with results from direct observations in 
{galaxy surveys at different redshifts}. We discuss the caveats {and} interpretations 
of our results, and 
{speculate} that the strong structural evolution 
found {in some previous studies for early-type galaxies could be 
explained partially by photometric changes rather than by intrinsic structural changes. 
}

\end{abstract}

\begin{keywords}
galaxies: elliptical and lenticular, cD -- galaxies: evolution -- galaxies: stellar contents -- galaxies: structure -- techniques: imaging spectroscopy
\end{keywords}


\section{Introduction}

Elliptical (E) galaxies occupy one extreme of the Hubble morphological classification as the most spheroid dominated systems. Their kinematic, photometric, and metallicity properties show them as the most evolved galaxies in the local Universe, and many pieces of evidence indicate that they suffered early violent dynamical processes that led to their spheroidal morphology (for recent reviews, see e.g., \citealp{Kormendy2016} and Chapter 10 in \citealp{Cimatti+2019}).
Studies based on observations from galaxy surveys at different redshifts have shown that most of the red, quiescent spheroid-dominated galaxies were much more compact in the past \citep[called ``red nuggets'';][]{vanDokkum+2008,Damjanov+2009} than the local spheroid-dominated galaxies, that is, early-type galaxies (ETGs). If red nuggets are the progenitors of massive present-day ETGs, then observations show that they increased their half-light or effective radii, \re, by a factor of $3$ from $z\sim 2$ to $z\sim0$ \citep[e.g.,][and more references therein]{Daddi+2005,Trujillo+2006,Trujillo+2007,vanDokkum+2010,Patel+2013,vanderWel+2014,Hill+2017,Mowla+2019}.

The above finds an explanation within the context of the two-phases formation scenario of ETGs \citep[][]{Oser+2010}. According to this scenario, in the early ($z\gtrsim 2$) {\it dissipative} phase driven by wet major mergers \citep[e.g.,][]{Hopkins+2008} and/or violent disc instability \citep[e.g.,][]{Barro+2013,DekelBurkert2014,Zolotov+2015,Tacchella+2016}, the proto-ellipticals are compact objects that form stars in situ in a very intense regime until this process is stopped abruptly by gas shock heated in massive halos, and/or rapid gas exhaustion and Supernova/AGN feedback. 
As the compact quiescent galaxy ages and reddens (red nugget), during the second {\it non-dissipative} phase ($z\lesssim 1-2$), more mass is assembled by the accretion of ex situ stars through merging with other galaxies, which are also mostly quiescent. Dry minor or intermediate mergers contribute little to the mass growth but promote substantial size growth (the accreted stars tend to be deposited mostly in the external regions), making spheroidal galaxies less compact and likely also less concentrated \citep[][among others]{Bezanson+2009,Naab+2009,Oser+2010,Trujillo+2011,Bluck+2012,Johansson+2012,Hilz+2013,Shankar+2013,vanDokkum+2015,Wellons+2016,Furlong+2017,Faisst+2017,Hill+2017,Genel+2018}. Dry mergers can also be major, though they are less frequent and relevant only to the most massive galaxies \citep[e.g.,][]{Bundy+2009,Lopez-SanJuan+2012,Rodriguez-Puebla+2017}.
These mergers increase substantially the mass, while the size increases approximately proportional to the mass increase  \citep[e.g.,][]{Nipoti+2003,Nipoti+2012,Johansson+2012,Hilz+2013}
in a such way that the shift in the mass--size relation is small, affecting in a lesser degree the compactness and concentration of the merged galaxies.
Although the second phase of ETG formation, driven by dry mergers, is a reliable explanation for the growth in size and the puffing-up of massive ETGs, there is still an intense debate as to whether or not this mechanism is enough to describe observational inferences \citep[e.g.,][see for a discussion \citealp{Zanisi+2021} and more references therein]{Newman+2012,Lopez-SanJuan+2012,Nipoti+2012,Sonnenfeld+2014,Man+2016,Frigo+2017}. An alternative or complementary mechanism suggested for the apparent strong growth in size of massive ETGs is quasar feedback, which removes huge amounts of cold gas from the central regions, inducing an expansion of the stellar distribution \citep[][but see \citealp{Trujillo+2011}]{Fan+2008}. 

Alternatively, the observational inferences of a rapid size, \re, evolution, of ETGs from cosmological surveys have been suggested to be apparent, at least partially, due to the so-called ``progenitor bias"  effect \citep{vanDokkum+1996}:
newly quenched galaxies, which are larger in size, enter the survey luminosity selection limit at lower redshifts, thus increasing the mean size distribution of the selected populations with respect to those at higher redshifts \citep[e.g.,][]{vanderWel+2009,Carollo+2013,Poggianti+2013,Shankar+2015}. 
This selection effect seems to explain most of \re\ increasing of quiescent galaxies with \ms$< 10^{11}$ \msun\ and for $z\lesssim 1$ \citep[e.g.,][]{Cassata+2013,Belli+2015,Carollo+2016,Fagioli+2016,Gargiulo+2017,Faisst+2017}, while for the most massive galaxies it may explain most of this increasing only at early epochs, $z> 1.5-2$ \citep[e.g.,][]{Belli+2015,Gargiulo+2017,Zanisi+2021}. 
Furthermore, there are other redshift-dependent selection effects and systematic uncertainties that may bias the inferences of the individual growth in size of galaxies \citep[see e.g.,][]{vanDokkum+2008,vanderWel+2009,Ribeiro+2016,Mosleh+2017,Genel+2018,Roy+2018,Whitney+2019}. 

Remarkably, {the change with $z$ of the color or mass-to-light ratio ($\Upsilon_\star$) gradients of quiescent galaxies has been also suggested to explain part of their claimed strong size evolution \citep[e.g.,][]{LaBarbera-Carvalho2009,Kennedy+2015,Ciocca+2017,Marian+2018,Suess+2019a,Suess+2019b}.} 
The latter authors, using data from the CANDELS survey, have shown that the radial variations in $\Upsilon_{\star}$ cause that the galaxy's light profile is different from its mass profile, in such a way that {\it the half-light radius is a biased tracer of galaxy mass distribution}. This is the main questions we aim to study in this paper by means of the fossil record {(or archaeological)} analysis of a large sample of local E galaxies.
 
\subsection{Galaxy evolution through the fossil record method}

The fossil record method {allows us to recover the mass, luminosity, and chemical enrichment histories of a stellar system out of the information encoded in its spectrum by using stellar population synthesis (SPS). The above method  applied to local galaxy surveys with Integral Field Spectroscopy (IFS) observations offers an alternative and complementary way of studying the global and spatially-resolved} evolution of galaxies \citep[see for a recent review][]{Sanchez2020}. In contrast to studies based on matching galaxy populations at different redshifts to infer about the evolution of galaxies \citep[e.g.,][]{vanDokkum+2010}, the fossil record inferences refer to the evolution of {\it individual} galaxies.
In this paper, we use the fossil record method for studying the radial evolution of the stellar populations for local E galaxies from the SDSS ``Mapping Nearby Galaxies at APO'' survey \citep[MaNGA,][]{Bundy+2015}. In particular, for a sample of ``dead and red'' E galaxies, referred also as Classical Ellipticals (CLEs), we study the differences between their half-light and half-mass radii over time.
Excluding Es with signatures of recent events of star formation (SF; e.g., blue, star-forming or recently quenched galaxies) allows us to avoid contamination in the spectra due to scarce but luminous young populations that affect the inferences of the structural evolution through the fossil record method.
The E--S0 galaxy separation is also important for this study to avoid the presence of discs, which are expected to have a different structural evolution than spheroids.

In a previous paper, \citet{Lacerna+2020}, we have found that the CLE galaxies accumulated 90\% (50\%) of their stellar masses between 5 and 7 (9 and 12) Gyr ago, with a downsizing trend in both stellar mass growth and SF quenching: the most massive CLEs tend to form stars earlier and to quench SF faster than the less massive CLE galaxies. 
As for radial trends, most of the CLEs have nearly flat radial mass-weighted age profiles but slightly negative when the luminosity-weighted ages are used. On the other hand, the stellar metallicity gradients are clearly negative. The spatially-resolved fossil record histories suggest that CLEs evolved {\it both by inside-out SF quenching and by inside-out mass growth}, being more pronounced the former than the latter.

This paper presents an update of the MaNGA CLE galaxy sample of \citet{Lacerna+2020}. By using the fossil record method, we calculate the 2D mass and light (for different rest-frame bands) spatially-resolved maps corresponding to stellar populations up to a given look-back time.
Our main goal is to study for the CLE archaeological progenitors, {\it how different evolve their radial light distribution from the radial mass distribution}. As in observational studies {from galaxy surveys at different redshifts}, to characterize these differences with global properties, we use the ratios of half-mass to half-light radii, as well as the ratios of light to mass concentrations, and of light to mass effective surface densities. 
{Thus, for the progenitors of present-day CLE galaxies, we calculate how much the photometric sizes deviate at different times from the intrinsic stellar mass sizes (or the photometric concentrations from the mass concentrations). } 
In a companion paper (Avila-Reese et al., in prep.), we will present the respective evolution of the $\Upsilon_{\star}$ (and color) gradients of the CLE galaxy progenitors and show that the change of these gradients over time drives the change of the differences between the mass and light-derived global structural properties found here.

The remaining of this paper is structured as follows. In Section \ref{sec:data}, we describe (i) the implementation of the fossil record method to reconstruct the archaeological structural parameters, and (ii) the sample selection criteria. In Section \ref{sec:results}, we present our results on the evolution of the differences between the radial mass and light distributions of the CLE progenitors, {where these difference are characterized by the following global quantities in mass and light: sizes (\S\S \ref{sec:radius-evol}), concentrations (\S\S \ref{sec:concentrations}), and mean surface densities (\S\S \ref{sec:surface-densities})}. In Section \ref{sec:comparisons}, we compare our results with a compilation of direct observations from galaxy samples and surveys at different redshifts. In Section \ref{sec:discussion}, we discuss the caveats of our study and how these caveats could impact our results. In addition, in this Section, we discuss the interpretation and implications of our results. Finally, in Section \ref{sec:conclusions}, we present our conclusions. 
In this paper, we adopt a standard $\Lambda$CDM cosmology, with $\Omega_{\rm m}=0.27$, $\Omega_\Lambda=0.73$, and  $H_0=71$ km s$^{-1}$ Mpc$^{-1}$.

\section{Methods, Sample, and galaxy characteristic radii} 
\label{sec:data}

\subsection{Stellar population analysis}
\label{S_ssp}

We make use of the \verb|Pipe3D| pipeline, which includes the FIT3D code for the SPS analysis \citep{Sanchez+2016_p21,Sanchez+2016_p171,Sanchez+2018_AGN}, to perform the spatially resolved stellar population study of the MaNGA data cubes. {\verb|Pipe3D| is an analysis tool for IFS data that has been used in many studies based on data from IFS surveys (e.g., CALIFA, MaNGA, and SAMI) and IFS instruments (e.g., MUSE); for some examples, see \citet[][]{Sanchez+2016_p21,Sanchez+2018_AGN,Cano-Diaz+2016,Cano-Diaz+2019,Ibarra-Medel+2016,Bellocchi+2019,Sanchez2020}.  }
Following, we briefly summarize how \verb|Pipe3D| works  \citep[for details and multiple tests, see][]{Sanchez+2016_p21,Sanchez+2016_p171,Ibarra-Medel+2016,Ibarra-Medel+2019}.  First, \verb|Pipe3D| performs an spatial segmentation or binning to achieve a homogeneous signal-to-noise (S/N) threshold level of $\sim 50$ on each segment. Then, \verb|Pipe3D| fits the stellar continuum on each segment by using the stellar population synthesis code \verb|FIT3D|.

{Here, we use \verb|FIT3D| with the implementation of  the GRANADA and MILES simple stellar population (SSP) libraries. The GRANADA and MILES (named gsd156) SSP library uses the \citet{Vazdekis:2010aa} templates, which starts at 63 Myrs, and it is complemented by younger stellar models from \citet{Gonzalez-Delgado+2005}, both matching in the same metallicity range. This library was introduced in \citet{CidFernandez+2014b} as GM (later named gsd156). The use of the \citet{Gonzalez-Delgado+2005} stellar models in the \citet{Vazdekis:2010aa} SSP library was intended to fill the gap between 1 Myr and 63 Myr. Both SSPs complement each other and cover 39 ages (14--0.001 Gyr)\footnote{In practice, we use results until 13 Gyr.} and 4 metallicities ($[Z/H]$=0.004, 0.008, 0.02, and 0.03) 
giving a suitable range to study the evolution of galaxies. The gsd156 library has been tested with multiple observations and simulations \citep[e.g.,][]{,CidFernandez+2013,CidFernandez+2014b,Sanchez+2016_p21,Sanchez+2016_p171,Ibarra-Medel+2019}. The gsd156 SSP library uses the \citet{Salpeter+55} initial mass function (IMF). Using this library,} \verb|FIT3D| performs the {\it non-parametric} SSP fitting to the spectra in the $3500\AA-7000\AA$ wavelength range.  In addition, \verb|FIT3D| models the effects of dust extinction in the SPS analysis using a \citet{Cardelli+1989} extinction law.

{After the spectral inversion through the SSP fitting with \verb|FIT3D|, \verb|Pipe3D| performs the analysis of the nebular emission lines, and creates a set of maps that contain the stellar population properties, the SSP outputs, the kinematics, the nebular emission and the nebular kinematics. These maps are processed by Pipe3D to undone the initial segmentation and generate 2D maps with the same number of spaxels as in the original data cube.
The stellar population model for each spaxel is estimated by re-scaling the best fitted model within each spatial segment to the continuum flux intensity in the corresponding spaxel.
In addition, in this work we post-process the SSP decomposition maps to obtain the spatially-resolved 2D maps across look-back time (\tlb), including the corresponding rest-frame spectra and photometric bands. We describe the latter in the next subsection. It should be said that several previous works have used the Pipe3D/FIT3D codes with other SSP libraries or introduced parametric SF histories, and they were compared with other codes using concrete observed or mock galaxies \citep[e.g.,][]{GonzalezDelgado+2015,Garcia-Benito+2017,Lopez-Fernandez+2018,Guidi+2018}. In most cases, the archaeological results indicated that they did not depend on these variations at a qualitative level.}

\subsubsection{Fossil record 2D map reconstruction}
\label{sec:fossil-record}

To generate the spatially-resolved 2D maps {with stellar spectral (in the rest-frame) and mass information across \tlb,} we reconstruct the archaeological evolution of the galaxy spectra. From the SSP decomposition, $f_{SSP}$, provided by \verb|FIT3D|, we know the fractional contribution in light of each SSP to the observed spectra in terms of the age ($t_i$) and metallicity ($Z_j$). In addition, we know from the gsd156 stellar library the values of the stellar mass loss ($m_{loss}$) and the mass-to-light ratio at any wave-length, $\Upsilon_{\star,\lambda}$, for each SSP, also in terms of the age and metallicity. Hence, the reconstructed spectrum {up to a given age ($>t$)} is given by:

\begin{equation}
\begin{split}
    F(\lambda,>t)=\sum_j\sum_i^{t\le t_{i}}f_{ssp}(Z_j,t_i)L_{ssp}(\lambda,Z_j,t_i-t)\\ \times L_V\times10^{0.4A_V}\times\frac{\Upsilon_{\star,\lambda}(Z_j,t_i)}{\Upsilon_{\star,\lambda}(Z_j,t_i-t)}\times\frac{m_{loss}(Z_j,t_i-t)}{m_{loss}(Z_j,t_i)}
\end{split}
\end{equation}
$L_V$ is the integrated luminosity over the Johnson's $V$ band and $A_V$ is the fitted extinction derived from the SSP decomposition. Therefore, the reconstructed spectra {up to the} age $t$ is the sum of all the SSP spectra with ages larger or equal to $t$, once we consider the changes on $m_{loss}$ and $\Upsilon_{\star,\lambda}$ across the ages. A SSP with age $t_i$ today would have an age $t_i-t$ at the epoch $t$. This reconstruction is applied for each spaxel in the data cube. To obtain the photometric luminosity {at rest frame} within an spaxel {up to given age} and at a given photometric band $photo$, $L(t)_{photo}$, we simply convolve the desired response function of the given photometric band to the reconstructed archaeological spectra. The archaeological reconstruction of the stellar mass at each spaxel {up to a given age} is the cumulative SF history taking into account the change of the stellar mass loss:

\begin{equation}
\begin{split}
    M_{\star}^{\rm P3D}(>t)=\sum_j\sum_i^{t\le t_{i}}f_{ssp}(Z_j,t_i)\times L_V\times10^{0.4A_V}\\ \times \Upsilon_{\star}(Z_j,t_i)\times \frac{m_{loss}(Z_j,t_i-t)}{m_{loss}(Z_j,t_i)}
\end{split}
\end{equation}
Finally, the mass-to-light ratio in the photometric band $photo$ {up to } the age $t$ is $\Upsilon(>t)_{\star,photo}\equiv$ \ms($>t$)$/L(>t)_{photo}$. 
In this work we use the SDSS $g,r$ and $i$ photometric bands.

\subsection{The sample of red and dead Ellipticals}
\label{sec:sample}

Our sample of E galaxies is based on the MaNGA data from the Sloan Digital Sky Survey (SDSS) Data Release 15 \citep[DR15;][]{DR15+2019}, which contains 4621 galaxies.
MaNGA is an IFS survey of $\sim 10,000$ galaxies in the redshift range of 0.01 $<$ $z$ $<$ 0.15 \citep[][]{Bundy+2015} at a spectral resolution of $R\sim 2000$ or $\sim$ 65 km s$^{-1}$ in the
wavelength range 3600--10300 \AA, and a median spatial resolution of 2.54 arcsec FWHM 
\citep[1.8 kpc at the median redshift of 0.037,][]{Drory+2015,Law+2015,Law+2016DRP}.
MaNGA targets are chosen from the NASA-Sloan Atlas catalogue (NSA; \citealp{Blanton+2005}) such that the distribution is nearly uniform in $\log$\ms\ \citep[][]{Wake+2017}. 

\subsubsection{Morphological classification of Elliptical galaxies}
\label{sec:morphology}

{We use here the morphological classification of MaNGA DR15 galaxies from \citet[][]{Vazquez-Mata+2021}, which is part of one of the official MaNGA Value Added Catalogs\footnote{ \url{https://www.sdss.org/dr16/data_access/value-added-catalogs/?vac_id=manga-visual-morphologies-from-sdss-and-desi-images}}, and it is reported in \citet[][]{DR17+2021}.
In \citet[][]{Vazquez-Mata+2021} the} morphological classification was carried out based on a visual inspection to a combination of (i) newly background-subtracted and gradient-removed\footnote{{Given a set of overlapping images, characterization of the overlap differences is key to determine how each image should be adjusted before combining them. \citet[][]{Vazquez-Mata+2021} took the approach of considering each image individually with respect to it neighbors. Specifically, they determine the areas of overlap between each image and its neighbors, and use the complete set of overlap pixels in a least-squares fit to determine how each image should be adjusted (e.g. what gradient and offset should be added/removed) to bring it ``best'' in line with its neighbors.}} SDSS $r$ band and $gri$ color images;  (ii) post-processed deep $r$ band images, $grz$ color images, and residual (after best model subtraction) images from the DESI Legacy Imaging Surveys \citep[][]{Dey+2019}. The above visual inspection allowed the authors to isolate E, S0 and S0a candidates as much as possible.

{Elliptical galaxies were identified by visually judging for the presence of a high central concentration of light with a gradual fall-off in brightness at all radii and outer regions having no sharp edges. In contrast, Lenticular galaxies were identified as those having a relatively prominent central light concentration with a sharp outer edge, that is, where the light drops off drastically showing a relatively flat profile from intermediate to outer radii. There are cases of prominent outer rings evidencing the presence of an extended disk in these galaxies, thus of S0 nature.}

{A possible source of confusion could be some poorly resolved face-on (large $b/a$ ratio) S0s at high $z$. However, for instance, \citet[][]{DominguezSanchez+2020} found evidence that (fast-rotating) E galaxies are not simply S0 galaxies that are viewed face-on. In the case of \citet[][]{Vazquez-Mata+2021}, the presence of (weak) rings or sharp edges in galaxies with large high $b/a$ ratios pre-selected as bulge dominated were used as criteria to separate S0--S0a from E types. 
On the other hand, E galaxies are not always composed of pure smooth spheroids.} The digital image processing to the DESI images in combination with the PSF-convolved DESI residual images, allowed the authors to identify E galaxies showing possible disc-like features embedded in the inner region of the spheroid. {When the disc-like feature is relatively bright and large, the galaxy was classified as S0. However, in most of the cases, the clear dominion of a structurally defined spheroid, led to classify them as E galaxies, but with inner discs (this type of Ellipticals are called Elliculars).} To this respect see, for example, the work by \citet{GrahamA+2019} emphasizing on the often overlooked continua of disc sizes in early-type galaxies. 

{In the early MaNGA morphological classification used in \citet{Lacerna+2020}, many of the E galaxies with an inner disc were classified as S0s. In the morphological re-evaluation of these galaxies in  \citet[][]{Vazquez-Mata+2021}, they were re-classified as Ellipticals (actually Elliculars), that is, the number of E galaxies increased with respect to \citet{Lacerna+2020}.}  These authors studied a sample of 340 MaNGA Ellipticals from DR15, but limited to $z\le$0.08. In the present paper that condition is relaxed, thus increasing our sample to an amount of 722 Ellipticals out to $z$ = 0.15 (504 out to $z=0.08$). 
Originally, 859 galaxies were classified as Ellipticals but 137 of them were discarded because of the presence of strong tidal features, large bright clumps or extended structures in the images within the MaNGA FoVs. 
{As discussed in \citet[][]{Vazquez-Mata+2021}, when comparing their morphological classification of MaNGA DR15 galaxies with the automatic classification by \citet[][]{DominguezSanchez+2021}, even including E galaxies with inner discs to the group of Ellipticals (as we do here), their fraction is lower than that of \citet[][]{DominguezSanchez+2021}. We suggest that the use of the DESI images and the visual detection of weak rings and sharp surface brightness edges, helps to separate better Lenticulars from Ellipticals.}

\subsubsection{Photometric properties}
The circularized effective radius in the $r$-band is calculated as \rer $= {\it a}_{\rm e,r}\times \sqrt{(b/a})_r$, where $a_{\rm e,r}$ is the half-light semi-major axis and $(b/a)_r$ is the minor-to-major axis ratio. 
We take these quantities from the MaNGA PyMorph DR15 photometric catalog \citep[][]{Fischer+2019}\footnote{\url{https://data.sdss.org/datamodel/files/MANGA_PHOTO/pymorph/PYMORPH_VER/manga-pymorph.html}} based on SDSS images. {These authors obtained the structural properties, including $a_{\rm e,r}$, by fitting parametric models to the 2D surface brightness profile convolved with the PSF of the galaxy image.}
We use their case of S\'ersic truncated model. We removed 39 Ellipticals in which the S\'ersic fit failed due to contamination, peculiarity, bad-image or bad model fit. 
We do not also consider other 25 E galaxies with \rer{} smaller than the  SDSS $r$-band PSF of 1.5 arcsec. In addition, we removed four galaxies with a stellar mass lower than the completeness limit given in \citet[][for details, see their appendix C]{Rodriguez-Puebla+2020}. 
The stellar mass is obtained from the NSA catalog \citep[][it uses a \citealp{Chabrier2003} IMF]{Blanton+2005}. 
Therefore, our pruned sample of Ellipticals consists of 654 galaxies. 

\subsubsection{Spectro-photometric selection of classical E galaxies}
\label{sec:spec-classification}

In this paper, we aim to study the evolution of the stellar mass and light radial surface distributions of the Ellipticals that are ``red and dead'', the CLE galaxies. By ``dead'' we understand that the galaxy has been in a quenching regime by a long time (see Appendix \ref{sec:quenching} for a disscusion on definitions of SF quenching), that is, it became passive or retired. The operational criteria to define red and retired galaxies were discussed in \citet[][]{Lacerna+2020}. A summary is as follows. 
Red and blue galaxies are separated using the criterion found by \citet{Lacerna+2014} in the $g-i$ color vs. \ms\ diagram  (colors were taken from the SDSS database with extinction corrected \textit{modelMag} magnitudes).  
As for the SF activity, we used the Pipe3D integral extinction-corrected H$\alpha$ equivalent width, EW(H$\alpha$), and the line ratio diagnostics in the BPT diagram \citep[][]{BPT1981}. 
Quiescent galaxies are defined as those with EW(H$\alpha$)$<3\AA$
\citep[]{Sanchez+2014, Cano-Diaz+2016, Cano-Diaz+2019, Sanchez2020};
most of our sample of E galaxies (92.8\%) obey this criterion. Finally, to assure a long-term quenching regime for our retired galaxies, we impose the criterion of an integral luminosity-weighted age larger than $4$ Gyr (see below).

The number of E galaxies obeying the three criteria mentioned above amounts to 537, that is, 82.1\% of our pruned sample of 654 Ellipticals are CLE galaxies.

The remaining 17.9\% of the MaNGA E galaxies have particular photometric and spectral line diagnostic features, which evidence them as different to the CLE galaxies, specially regarding their late evolution \citep[for details, see][]{Lacerna+2020}. 
Briefly, the remaining E galaxies in our sample are as follows.
A fraction of 6.9\% of the E galaxies have either EW(H$\alpha$)$<3\AA$ and luminosity-weighted ages $\leq4$ Gyr or $3\leq$ EW(H$\alpha$)/$\AA<6$ lying above the \citet{Kewley+2001} relation in the BPT diagram. Following \citet{McIntosh+2014}, we identify these galaxies as Recently Quenched Ellipticals (RQE). As discussed in \citet{Lacerna+2020}, RQEs have different properties and evolutionary paths with respect to CLEs. The most massive RQEs could be the result of late ($z<0.3-0.5$) major or intermediate mergers, while the less massive RQEs could be associated to the accretion of gas-rich satellite(s) or gas infall since $z\lesssim 0.8$.  
A small fraction (2.6\%) of the E galaxies are blue and star-forming (BSF),
which probably are E galaxies that rejuvenated recently by cosmic gas infall. Galaxies are considered as star-forming when EW(H$\alpha$)$>6\AA$ and the galaxy lies below the \citet{Kewley+2001} relation. 
The remaining fraction of MaNGA Ellipticals (8.4\%) are either retired but blue, red but star-forming, undetermined or with AGN.\footnote{We note that the fractions of CLEs, RQEs, and BSF Ellipticals changed with respect to reported in \citet[][it increases in the case of CLEs, whereas it decreases for RQEs and BSF]{Lacerna+2020} mostly because we do not establish here an upper limit of $z = 0.08$ as in that paper. The current sample considers galaxies out to $z \approx 0.15$, which includes a higher fraction of red, retired galaxies than in the case of limiting to $z=0.08$.}.

\begin{figure}
	\includegraphics[width=\columnwidth]{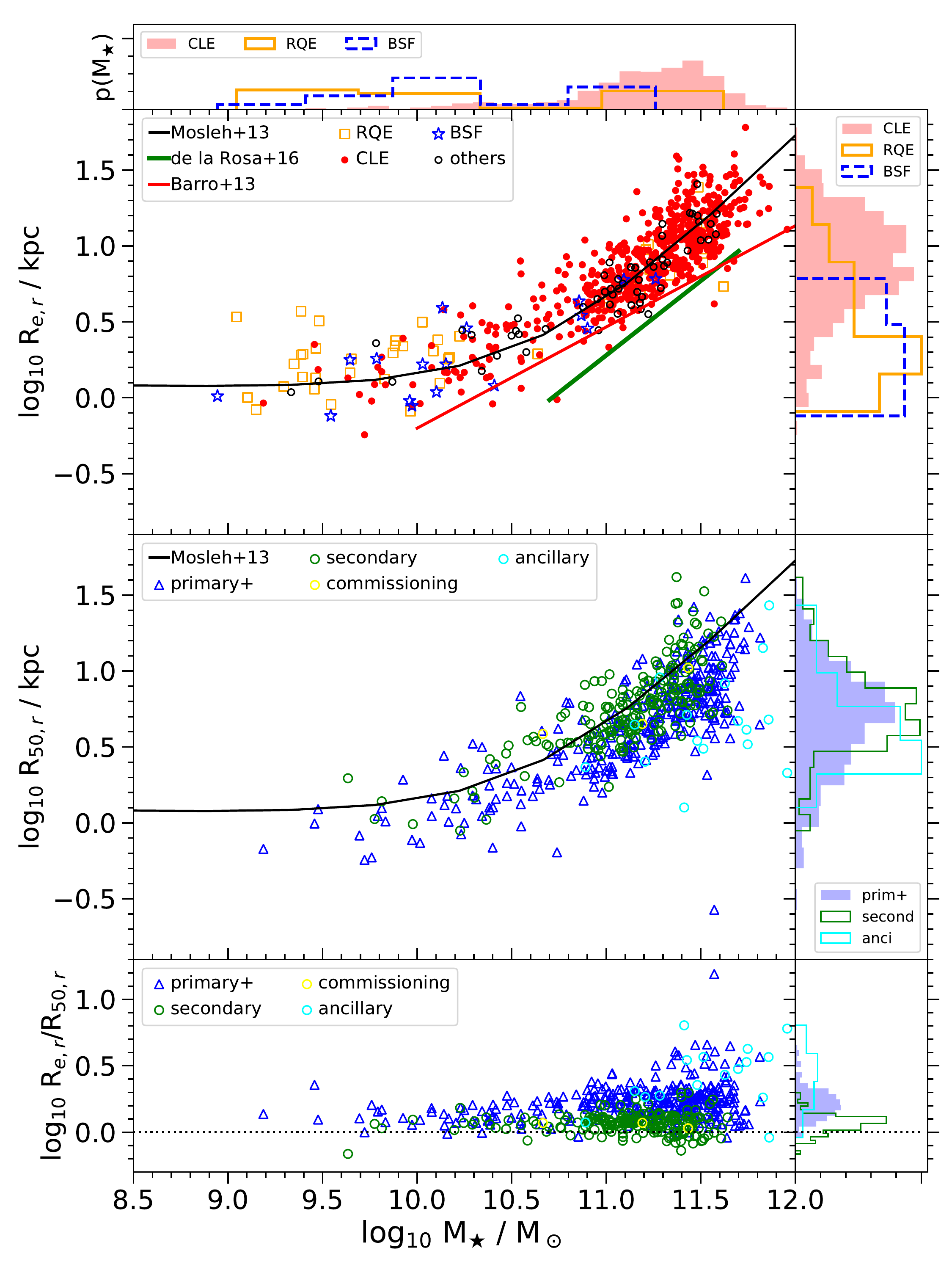}
    \caption{Mass--size relation of MaNGA Ellipticals. 
    {\it Upper main panel:} Stellar mass vs. $r-$band effective circularized radius, \rer. The CLE galaxies are shown in red filled circles. We also show the RQE galaxies (orange open squares), BSF E galaxies (blue open stars), and other Es (black open circles). 
    The black line is the fit to the \ms--\rer\ relation for nearby early-type galaxies from \citet{Mosleh+2013}. 
    The red and green lines are the criteria of compact spheroids from \citet{Barro+2013} and compact cores from \citet{deLaRosa+2016}, respectively.
    Top and upper right sub-panels show the normalized density distributions of \ms\ and \rer, respectively, for CLE (red solid histogram), RQE (orange open histogram), and BSF (blue dashed histogram) galaxies. 
    {\it Middle main panel:} Stellar mass vs. \rlightr{} of only CLEs, where \rlightr{} is as calculated by Pipe3D from the MaNGA datacubes, within their limited FoVs. Blue triangles are for the Primary+ sample, whereas green, yellow, and cyan circles for Secondary, commissioning and ancillary samples, respectively.
    The middle right sub-panel shows the respective normalized density distributions of \rlightr{}.
    {\it Lower main panel:} Differences between \rer{} and \rlightr{} of CLEs. The symbols are the same as in the middle main panel. 
    The lower right sub-panel shows the normalized density distributions of the differences in sizes with the same symbols as in the middle right sub-panel.
    The integral of each histogram in the sub-panels sums to unity.
    }
    \label{fig:mass-size-relation}
\end{figure}

\begin{figure*}
	\includegraphics[width=2\columnwidth]{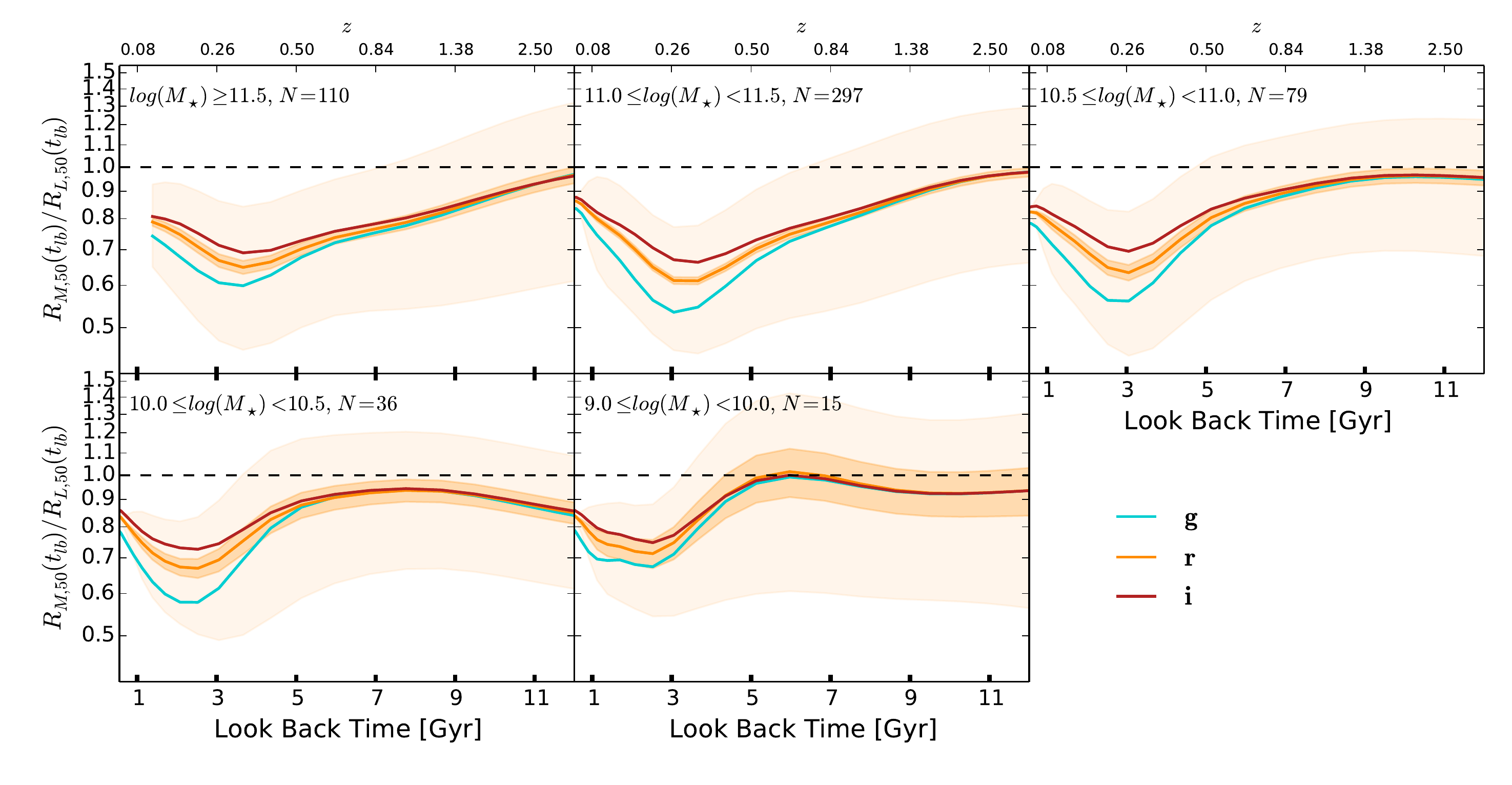}
    \caption{Evolution of the ratio between the half-mass and half-light radii of the MaNGA CLE archaeological progenitors. Each panel shows the running medians in different bins of \ms\ at the observation time for the $g,$ $r,$ and $i$ bands (cyan, orange, and red lines, respectively). The dark and light orange shaded regions correspond to the standard error of the median and the associated first and third quartiles for the $r$ band, respectively. The \ms\ range and the number of galaxies in each bin are indicated in the top of the respective panels.}
    \label{fig:radial-ratio-evol}
\end{figure*}

\subsection{Half-light radii from SDSS photometry and MaNGA data cubes}
\label{sec:radii-datacubes}

In the upper main panel of Figure \ref{fig:mass-size-relation}, we present our MaNGA sample of CLE galaxies in the \rer--\ms\ diagram (red circles). As mentioned above, to calculate the circularized effective radii we use the MaNGA PyMorph photometric catalog based on SDSS images \citep[][]{Fischer+2019}. 
We also include with different symbols the different classes of E galaxies as discussed above. The solid line corresponds to the best-fit relation for early-type galaxies as determined by \citet{Mosleh+2013} from a robust photometric analysis of SDSS galaxies. This relation also describes very well the MaNGA sample of E galaxies. The size is roughly constant with mass up to \ms $\sim$ 10$^{10.5}$ \msun{}, and it increases rapidly for more massive galaxies. We note that at low masses the RQE galaxies tend to be slightly above the  \citet{Mosleh+2013} relation.
We show the criterion that defines (extreme) compact early-type galaxies at $z\sim0$ following the criterion for fossil red nuggets given in \citet[][red line]{Barro+2013}. 
We also show the local compact core criterion (green line) from the compilation by \citet[][see their Figure 3, where they use the \citealp{vanDokkum+2015} criterion]{deLaRosa+2016}.  
Only 12 CLE galaxies can be considered as fossil red nuggets, i.e., below the red line.  If a more restrictive criterion for red nuggets is used, e.g., the one by \citet[][]{vanDokkum+2015}, then this number is smaller. 

In the middle main panel of Figure \ref{fig:mass-size-relation}, the MaNGA sample of only CLEs (red filled circles in the upper panel) is shown again in the size--mass diagram but this time using the $r$-band circularized half-light radius, \rlightr,  derived with Pipe3D from the MaNGA data cubes.
We first calculate the half-light semi-major radius, $a_{50,r}$, using the growth curve in elliptical isophotes along the semi-major axis from the 2D $r-$ band luminosity map and correcting it by the PSF broadening size:  $a_{50,r}= \sqrt{a_{50,r,{\rm meas}}^2 - a_{\rm PSF}^2}$, 
where $a_{50,r,{\rm meas}}$ is the {measured major semi-axis value at which half of the flux (within the FoV) is attained. This is a common way to roughly correct by the PSF size the half-light radius obtained non-parametrically from the growth curve \citep[e.g.,][]{Mosleh+2013}}. Then, as before, we define the circularized value as \rlightr$= a_{50,r}\times \sqrt{(b/a})_r$. {The same procedure applies for other bands as well as for the surface density mass.}
In Figure \ref{fig:mass-size-relation}, the symbols correspond to different spatial coverage in MaNGA. The FoV of MaNGA galaxies is such that it should cover up to $\sim1.5$\rer\ of the galaxy in the Primary+ sample (blue triangles) and up to $\sim2.5$\rer\ in the Secondary sample (green circles). The fraction of the Primary+ and Secondary galaxies for the CLE sample are $60.1\%$ and $36.3\%$, respectively (the remaining 3.0\% and 0.6\% correspond to ancillary and commissioning galaxies, respectively). Note that we use the subscript $e$ for the effective radii from the SDSS photometry, while the subscript 50 for the half-light radii {obtained non-parametrically} from the Pipe3D luminosity maps within the FoVs of the MaNGA observations. Due to the limited MaNGA FoV's, specially for the Primary+ sample, the latter are smaller than the former as seen in the lower panels. The median differences in the sizes are 0.22 dex for the Primary+ sample and only 0.072 dex for the Secondary sample. 
{The main expected impact of the limited aperture of MaNGA observations on our results is that the archaeological evolution of the outermost regions of the CLEs, where inside-out growth by minor mergers could be important \citep[see e.g.,][]{Mosleh+2020}, at least for the more massive, is not captured. Therefore, our inferences about late size growth and \rmass-to-\rlight\ evolution of these galaxies could be underestimated. }

\section{Results}
\label{sec:results}

Our goal here is to explore how different is the evolution of the radial stellar light distribution from the stellar mass one for our MaNGA CLE galaxies. To quantify the differences with global observable properties, we use the ratios of light to mass corresponding to three properties: radius (\S\S \ref{sec:radius-evol}), concentration (\ref{sec:concentrations}), and effective surface density (\ref{sec:surface-densities}). To calculate these properties, we use the 2D light and mass maps at different cosmic times, within the corresponding FoV of each observed galaxy. 
To define cosmic times, we go from the SSP age distribution of each galaxy to a common cosmic look-back time, \tlb, such that \tlb$= t_{\rm lb,obs} + t$, where $t_{\rm lb,obs}$ is the look-back time corresponding to the observational redshift of a given galaxy and $t$ is the given SSP age.
We use \mspipe(\tlb) to refer to the cumulative stellar masses calculated with Pipe3D within the MANGA FoV at different look-backs times, taking into account the stellar mass loss.


\subsection{Evolution of the ratio between half-light and half-mass radii}
\label{sec:radius-evol}

{We measure the half-light and half-mass semi-major radii as the scales where half of the total luminosity and mass are attained, respectively, within the FoV of each galaxy. The respective growth curves along the semi-major axis of elliptical annuli are used.}
We circularize these radii and correct by the PSF size (see \S\S \ref{sec:radii-datacubes}) to calculate \rlight\ and \rmass\ at each \tlb.
In Figure \ref{fig:radial-ratio-evol}, {where the main result of the present work is shown,} we plot the running median  in different stellar mass bins of the \rmass/\rlight\ evolutionary tracks of the archaeological progenitors of our CLE galaxies.  
The  \rmass-to-\rlight\ ratio tell us how compact a galaxy is in mass with respect to light (in different bands). The galaxies were labeled by their masses from the NSA catalog and grouped into five \ms\ bins indicated within the panels of Figure \ref{fig:radial-ratio-evol}. We present results in three \textit{rest-frame} bands, $g, r,$ and $i$. For the $r$ band, we also show the standard error of the median\footnote{We calculate the standard error of the median as $\sigma_{\rm med}=1.253 \sigma/\sqrt{N}$, where $\sigma$ is the
standard error and $N$ is the number of elements in the given bin. 
This formula is valid for a normal distribution. The distributions of the \rmass-to-\rlight\ ratios at different epochs are not normal but the deviations from it are small in most of the cases. Therefore, the above formula is a good approximation.} 
and the associated first and third quartiles in dark and light shaded regions, respectively.
The corresponding errors of the median and quartiles in the $g$ and $i$ bands are in general slightly larger and smaller than in the $r$ band, respectively. We do not show them to avoid over-plotting. 

As observed in Figure \ref{fig:radial-ratio-evol}, the evolution of the \rmass-to-\rlight\ ratio has some dependence on the final masses of the CLE galaxies. At early epochs, \tlb{} $\gtrsim$ 10 Gyr, 
the \rmass-to-\rlight\ ratio has a large dispersion, with median values around 1, and the compactness in different bands is roughly the same on average. 
At lower look-back times (lower $z$), the larger the \ms, the earlier the \rmass/\rlight\ ratio, on average, starts to decrease for all bands. The decreasing of the \rmass/\rlight\ ratios is more pronounced in the bluer bands. 
Within the large scatter, the average of the \rmass-to-\rlight\ ratios in the $r$ band reaches a minimum at \tlb$\sim 2$ Gyr ($z\sim 0.2$) for galaxies in the $9.0\le\log$(\ms/\msun)$<10.0$ bin, and at \tlb$\sim 4$ Gyr ($z\sim 0.35$) for those in the $\log$(\ms/\msun)$\ge 11.5$ bin.

The progenitors of less massive CLEs attain smaller differences between \rmass\ and \rlight\ on average, that is, their \rmass-to-\rlight\ ratios are higher at the minimum. For the massive CLEs, $\log$(\ms/\msun)$>10.5$, the minimum ratio in the $r$-band is $0.60-0.65$ on average.  After the minimum, the \rmass-to-\rlight\ ratios increase towards the present day, and the scatter significantly decreases.
For CLEs less massive than $\sim 10^{10}$ \msun, the \rmass/\rlight\ evolutionary tracks are scattered around the value of 1, and even with a tendency, in many cases, to values larger than 1, at \tlb$\gtrsim5$ Gyr.
{That \rlight<\rmass\ implies that the  light radial distribution (young populations) is more compact than that of stellar mass (older populations), which suggests that the galaxy is in a phase of strong SF bursts in the center \citep[e.g.,][]{Zolotov+2015,Tacchella+2016,Barro+2017}. Note that for more massive CLEs, at least for a fraction of them, this phase seems to have occurred earlier than for the less massive ones. }
At \tlb$\lesssim 5$Gyr, the mean \rmass-to-\rlight\ ratio of the less massive CLEs decreases and then increases towards the present day. 
This is the same U-shaped trend seen in the more massive CLEs but more compressed in time and amplitude.

In general, we observe that the less massive CLE galaxies follow the qualitative trends of the more massive ones with regard to the evolution of the \rmass-to-\rlight\ ratios but shifted to later epochs and with less pronounced trends.  
According to Figure \ref{fig:radial-ratio-evol}, on average, the more massive the CLE galaxy, the sooner its archaeological progenitor enters to the phase of decreasing systematically its \rmass-to-\rlight\ ratio. For the progenitors of CLEs more massive than $\sim 10^{11}$ \msun, this early phase happened at $z\sim2.5$ or higher, while for the least massive CLEs, this happens at $z\sim 0.5$, on average. 

The systematical decrease of the \rmass-to-\rlight\ ratio as \tlb\ is smaller seen in Figure \ref{fig:radial-ratio-evol} means that the galaxy becomes internally less compact (or externally more extended) in light  than in stellar mass, being more pronounced this trend as bluer the band is. {The above could be because the galaxy undergoes a gradual quenching of SF from the inside out and/or 
because the archaeological SF history is more prolonged in the outer regions of the galaxy, probably due to a lower SF efficiency there or to the late inside-out mass aggregation of younger populations by minor mergers \citep[ex situ stars are more common in the outermost regions, $\gtrsim 1.5-2$\re, though important fractions can be found also at inner radii, specially in the most massive galaxies, see e.g.,][]{Rodriguez-Gomez+2016}. 
Recall that the half-light and half-mass radii are associated to younger and older stellar populations, respectively, particularly in the rest-frame $g$ band.
In this sense, the \rmass-to-\rlight\ ratio quantifies the strength of the differences in stellar population ages along the galaxy, and is associated to the $\Upsilon_\star$ and color radial gradients \citep[see e.g.,][]{Chan+2016,Chan+2018,Suess+2019a}. 
On the other hand, the upturn and the subsequent trend of increasing the \rlight-to-\rmass\ ratio at late epochs seen in Figure \ref{fig:radial-ratio-evol} implies that after some epoch the light radial distribution approximates to that of the mass. 
If the light distribution is more extended than the mass one (\rlight$>$\rmass), but at some time the SF rate, SFR, strongly declines including the outer regions (global quenching), then as the stellar populations age passively, without significant structural changes, the shape of the stellar light profile in the optical bands (which traces relatively young populations), specially in the outer regions, will tend to the same shape of the profile of old populations (stellar mass profile), and then \rlight\ tends to \rmass. }
 In \S\S \ref{sec:interpretations} the above-mentioned interpretations for the \rmass/\rlight\ evolution of the CLE archaeological progenitors shown in Figure \ref{fig:radial-ratio-evol} are discussed in more detail.

{We note that inferences of the individual radial distributions of stellar light and mass at different ages from the fossil record method could be misleading due to the effects of radial mixing  (by stellar migration or mergers) in the observed stellar populations. Mixing tends to flatten stellar population radial gradients (see for a discussion \S\S \ref{sec:mergers}). As a result, the inferred half-mass (or half-light) radii of the archaeological progenitors at different ages tends to equalize,  possibly giving the impression of a smaller-than-actual size evolution. 
This effect is less important for the ratios of \rmass\ to \rlight\ because older and younger stellar populations, those that are traced by mass and light, respectively, shifted radially in roughly the same way (actually, older populations are susceptible to more shift than those that form later simply because the mixing mechanisms act longer).  
In other words, if the mixing effects lead us to estimate an archaeological growth of \rmass\ less than the true one, approximately the same applies to \rlight, specially in the more infrared bands. Therefore, while the archaeological inference of the \rmass\ and \rlight\ evolution could differ from the true evolution due to the possible radial mixing of stellar populations in the observed galaxy, their ratio is less affected by this effect. 

Bearing in mind the  above-mentioned caveat, in Figure \ref{fig:R-evolution} in Appendix \ref{appendix:R-evolution}, we present the median tracks of \rmass\ and \rlight\ as a function of \tlb{} in the same mass bins and for the same three photometric bands as in Figure \ref{fig:radial-ratio-evol}. The fossil record analysis for the CLE galaxies shows that, on average, \rmass\ almost does not change with \tlb{}. If any, for the more massive ones, \rmass\ is slightly smaller at larger look-back times. Therefore, is the evolution of the \rlight{} radii which actually drives the behaviors of the \rmass-to-\rlight\ ratios observed in Figure \ref{fig:radial-ratio-evol}. 
}


\begin{figure}
	\includegraphics[width=\columnwidth]{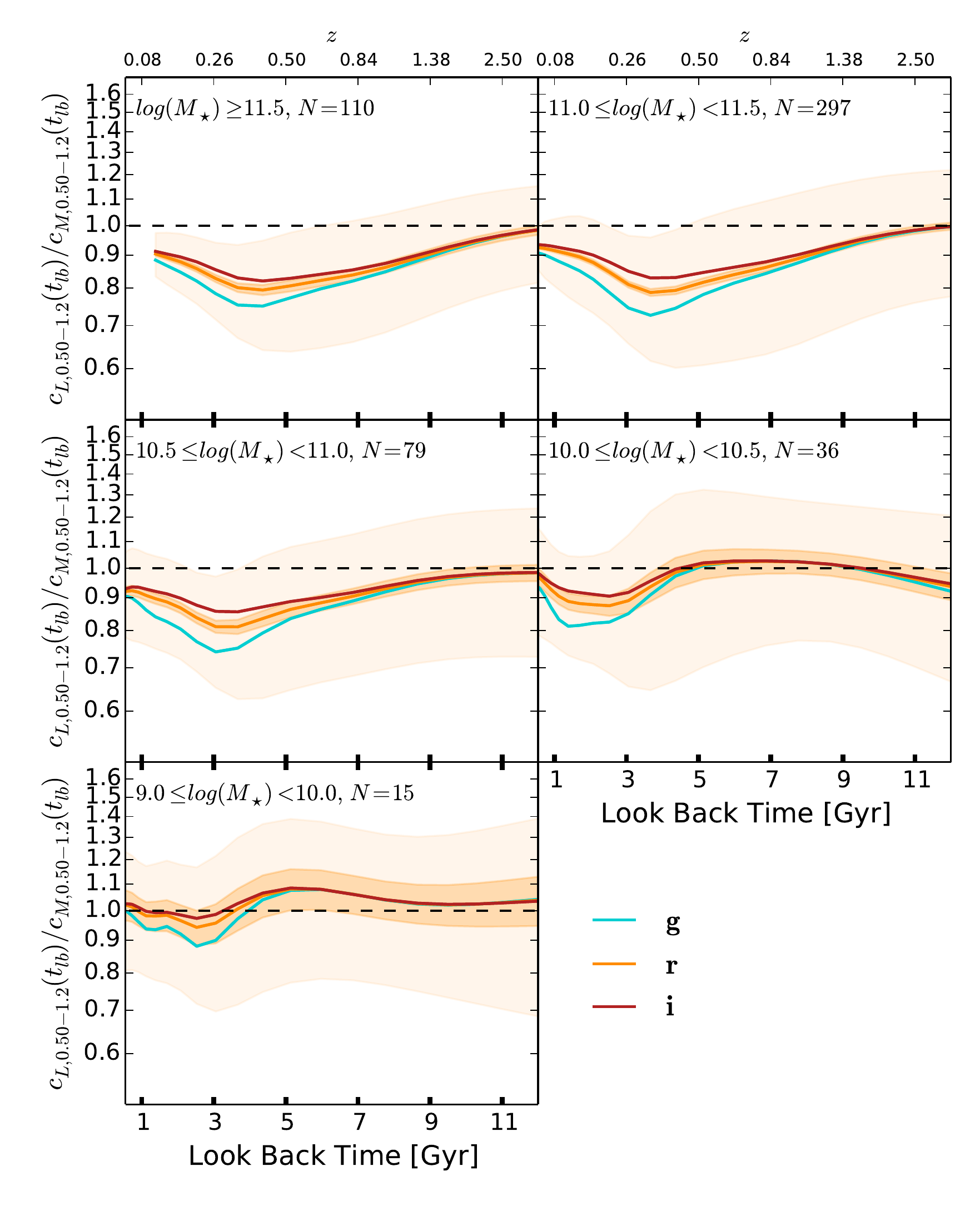} 
    \caption{Evolution of the ratio between stellar light and mass inner concentrations. The line and color codes are as in Figure \ref{fig:radial-ratio-evol}. 
    The trends are similar to those of the \rmass-to-\rlight\ ratio but the differences in the case of concentration, both in the stellar light-to-mass ratios and among the different photometric bands, are smaller than for the  size ratios. }
    \label{fig:concentration-evol}
\end{figure}

\begin{figure}
	\includegraphics[width=\columnwidth]{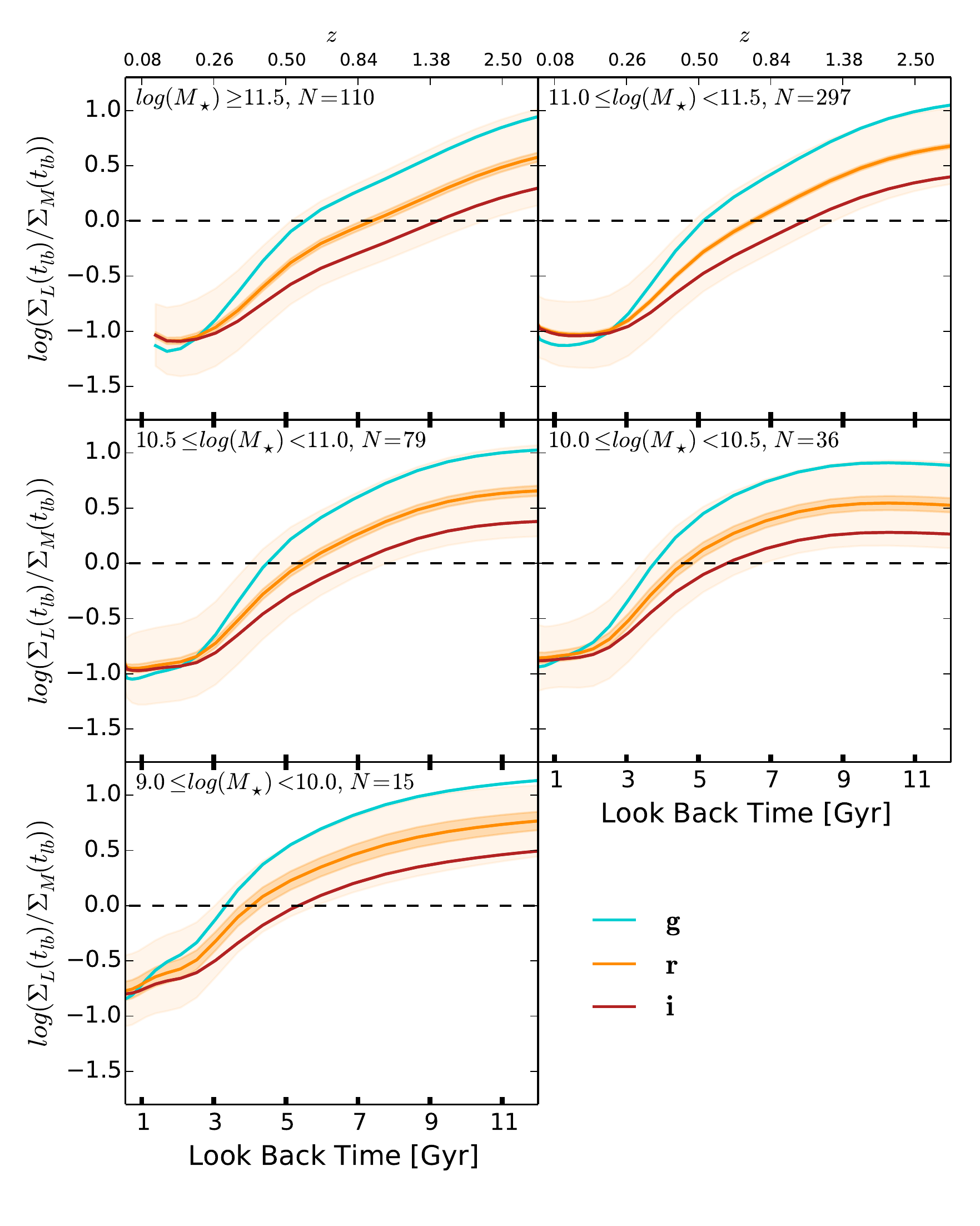} 
    \caption{Evolution of the ratio between effective surface brightness and effective surface density in the $g,$ $r,$ and $i$ bands.  The line and color codes are as in Figure \ref{fig:radial-ratio-evol}. 
    }
    \label{fig:sigmas}
\end{figure}

\begin{figure}
	\includegraphics[width=\columnwidth]{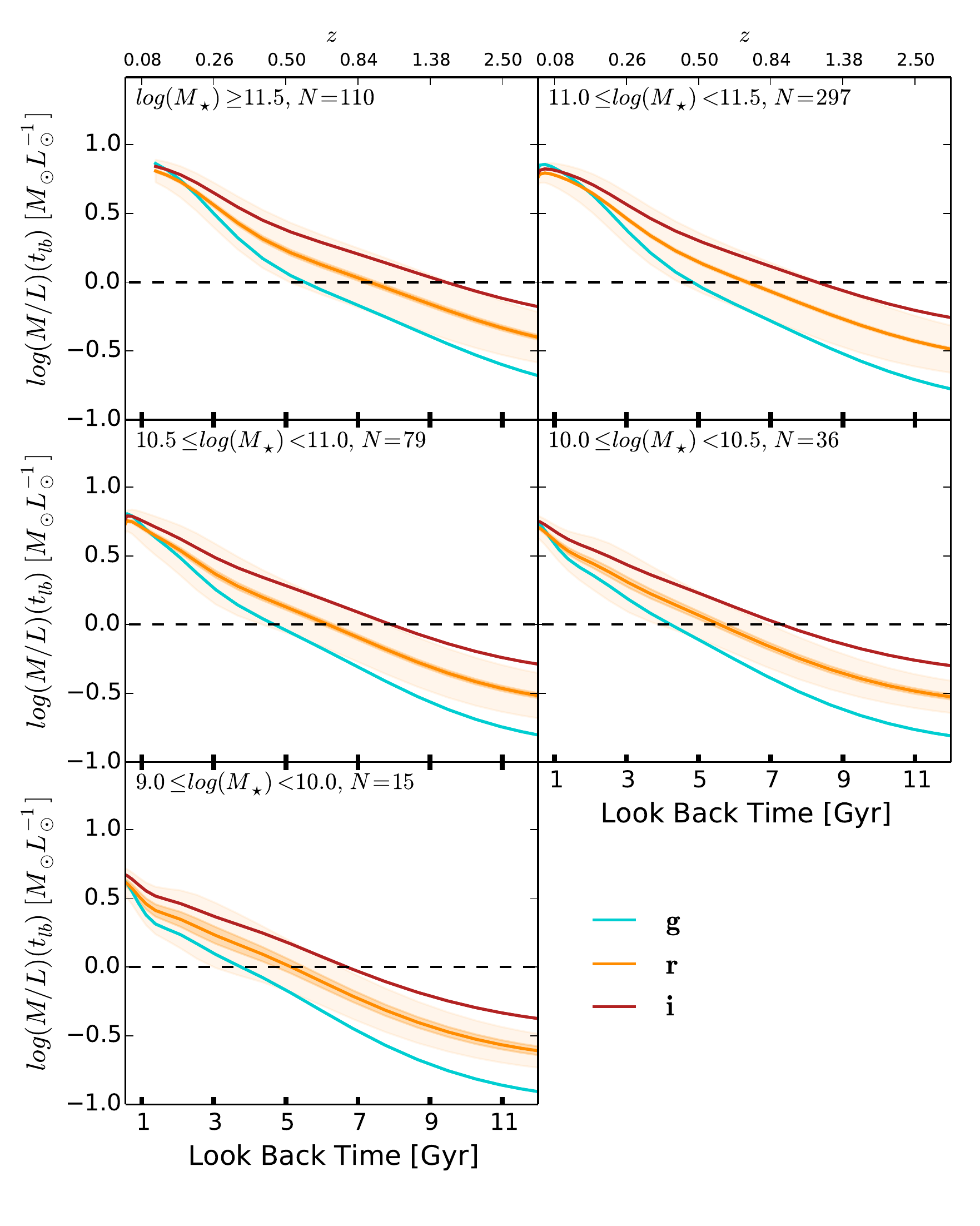}
    \caption{Evolution of the global mass-to-luminosity, $\Upsilon_\star$, in  the $g,$ $r,$ and $i$ bands.  The line and color codes are as in Figure \ref{fig:radial-ratio-evol}. 
    }
    \label{fig:mass-light-evol}
\end{figure}


\subsection{Evolution of the ratio between stellar light and stellar mass concentrations}
\label{sec:concentrations}

The \rmass-to-\rlight\ ratio studied in \S\S \ref{sec:radius-evol} tell us about how {\it compact} the galaxies in stellar mass are with respect to luminosity in different bands. Differences between the galaxy surface mass density and brightness radial distributions could be due not only to differences between the respective characteristic sizes, but also to differences between the shapes of these radial distributions, in particular to how peaked is the inner mass distribution with respect to the luminosity one. The latter is related to the {\it concentration} of the mass and light radial profiles. In other words, compactness and concentration are different concepts \citep[see e.g.,][]{Vulcani+2014,Kennedy+2015,Andreon2020}. The concentration can be evaluated parametrically, for example, through the S\'ersic index of the surface mass/brightness profiles fitted to a S\'ersic function, or non-parametrically, by using concentration indexes that measure the amount of mass and luminosity accumulated in two characteristic radii. Here, bearing in mind the limitations of the MaNGA footprint, we define the inner concentration in mass or luminosity up to a given look-back time as:
\begin{equation}
    \conc (<t_{\rm lb}) = \frac{\langle X\rangle_{R\le 0.5R_e}(<t_{\rm lb})}{\langle X\rangle_{R\le1.2R_e}(<t_{\rm lb})},
    \label{eq:conc}
\end{equation}
where $\langle X\rangle_{\rm R\le kR_e}$ 
is the mean (per unit of area) stellar mass or luminosity, $X=M_\star^{\rm P3D}$ or $L^{\rm P3D}$, contained in all spaxels within $k$\re\ using the SDSS observed \re\ (see \S\S \ref{sec:sample}).
This measure ensures that the extent of the mass and luminosity radial profiles at any epoch are scaled to a physical size, the effective radius of the observed galaxy, \re.  Our concentration parameter is similar to that one introduced by \citet[][]{Peterken+2020} with the difference that these authors measured the stellar mass concentration corresponding to stellar populations {\it of a given age interval},  while in our case, \conc\ refers to all stellar populations {\it up to} a given age. On the other hand, note that our interest here is in the evolution of the ratio between the stellar light and mass concentrations, rather than in the evolution of the concentrations per se.  
Values of $C_{0.5-1.2}^L/C_{0.5-1.2}^M$ close to 1 imply that the shapes of the inner radial distribution in mass and luminosity are similar, while $C_{0.5-1.2}^L/C_{0.5-1.2}^M<1$ means that the shape of the stellar mass distribution is ``peaker'' in its center than the shape of the stellar light distribution.

Figure \ref{fig:concentration-evol} shows the running medians, the standard errors of the median, and the associated first and third quartiles of the ratios between stellar light and  mass concentrations, $C_{0.5-1.2}^L/C_{0.5-1.2}^M$, in five \ms\ bins using the same line and color codes as in Figure  \ref{fig:radial-ratio-evol}. 
The trends are qualitatively similar to the \rmass-to-\rlight\ ratio in Figure \ref{fig:radial-ratio-evol}. If anything, the differences in the inner concentrations between mass and luminosity are smaller than the respective differences in size; the differences between different bands in concentration are also smaller than in size. In particular, at the smallest look-back times, the mass and luminosity concentrations of the CLE galaxies are very similar, closer to each other for all the bands than the differences between the half-mass and half-light radii.

\subsection{Evolution of the ratio between stellar light and mass effective surface densities}
\label{sec:surface-densities}

Besides the changes in the \rmass-to-\rlight\ and $C_{0.5-1.2}^L$-to-$C_{0.5-1.2}^M$ ratios reported in Figures \ref{fig:radial-ratio-evol} and \ref{fig:concentration-evol}, respectively, we can also explore the evolution of the ratio between the effective surface brightness, $\langle\Sigma\rangle_{\rm 50L}=L^{\rm P3D}(<R_{\rm 50L})/\pi R_{\rm 50L}^2$, and  effective stellar surface density, $\langle\Sigma\rangle_{\rm 50M}=M_\star^{\rm P3D}(<R_{\rm 50M})/\pi R_{\rm 50M}^2$. 
We calculate $\langle\Sigma\rangle_{\rm 50L}$ and $\langle\Sigma\rangle_{\rm 50M}$ up to a given look-back time from the photometric and stellar mass 2D maps described in \S\S \ref{S_ssp}, and obtain their ratio:
\begin{equation}
    \frac{\langle\Sigma\rangle_{\rm 50L}}{\langle\Sigma\rangle_{\rm 50M}} = \frac{L^{\rm P3D}(<R_{\rm 50L})/\pi R_{\rm 50L}^2}{M_\star^{\rm P3D}(<R_{\rm 50M})/\pi R_{\rm 50M}^2} \approx \frac{1}{\Upsilon_\star} \left(\frac{R_{\rm 50M}}{ R_{\rm 50L}}\right)^2.
    \label{eq:surf-dens}
\end{equation}
The approximation in the last term is because $\Upsilon_\star$ refers to the mass-to-light ratio integrated within the whole FoV, while we calculate the luminosity and stellar mass within their respective half-light and half-mass radii; the differences between $\Upsilon_\star$ measured within the FoV and within the half-light and half-mass radii are small and roughly the same at all epochs. 
Figure \ref{fig:sigmas} shows the running medians of $\langle\Sigma\rangle_{\rm 50L}/\langle\Sigma\rangle_{\rm 50M}$ as a function of \tlb\ in five present-day \ms\ bins for the $g,$ $r,$ and $i$ bands. The dark and light shaded bands are the standard error of the median and the scatter (first and third quartiles) for the $r$ band. 
The effective surface brightness of CLE archaeological progenitors dramatically decreases over time with respect to the effective surface density, more in the bluer rest-frame bands. Also, these trends are amplified as \ms\ is larger. At late epochs (low redshifts), the  $\langle\Sigma\rangle_{\rm 50L}/\langle\Sigma\rangle_{\rm 50M}$ ratio tends to decrease slower than in the past. According to Eq. (\ref{eq:surf-dens}), the above results are the combination of the changes over time of $\Upsilon_\star$ and the \rmass-to-\rlight\ ratio. The latter was shown in Figure \ref{fig:radial-ratio-evol}, while the former is shown in Figure \ref{fig:mass-light-evol}.
Over time, $\Upsilon_{\star}$ increases significantly in the three bands and for all masses up to the present-day values of $\Upsilon_{\star}\approx 5-7$.

{Based on our results, if photometric observations of galaxies at different redshifts expected to end as CLEs show a strong decrease in their surface brightness over time, then their actual decrease in surface density should be much more moderate. That is, any estimation of the evolution of a characteristic surface density for the progenitors of E galaxies (from photometric observations at different redshifts) must take into account the evolution in $\Upsilon_{\star}$ and \rmass/\rlight\ (a proxy of the $\Upsilon_{\star}$ gradient) that these galaxies have. In particular, as showed in Figure \ref{fig:mass-light-evol}, the former strongly increases with time. 
}


\begin{figure*}
	\includegraphics[width=2\columnwidth]{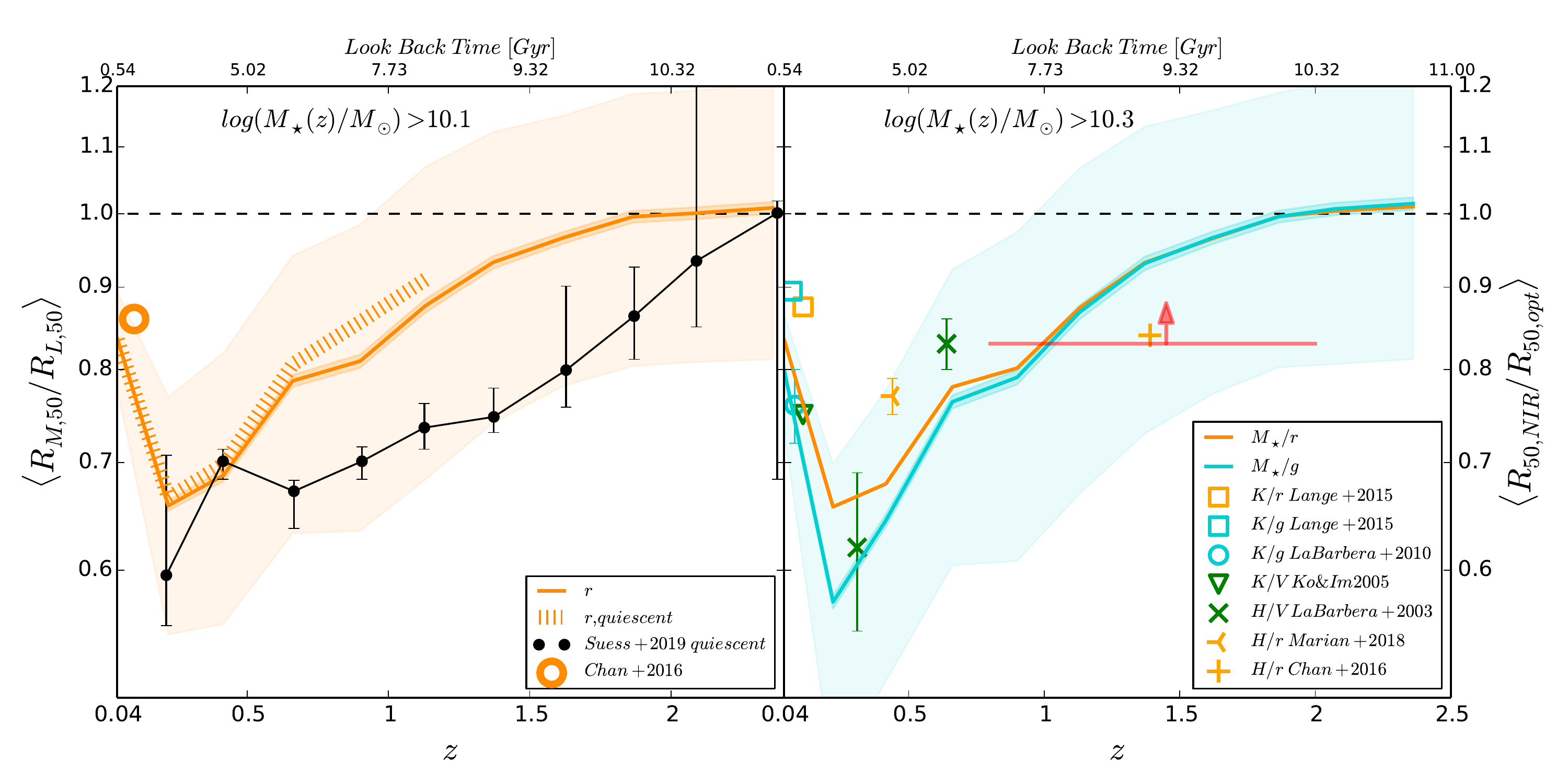}
    \caption{Comparison of the mass-to-light ratios in size as a function of $z$  or \tlb\ from our fossil record inferences and direct observations from surveys at different $z'$s. {\it Left panel:} Median of the \rmass-to-\rlightr\ ratios (orange solid line) and the associated first to third quartiles (light shaded region) for our CLE galaxy progenitors with $\log$(\ms($z$)/\msun)$>10.1$ \msun. Black dots with error bars show the results for quiescent galaxies from the CANDELS survey \citep[][]{Suess+2019a,Suess+2019b}. 
    The vertical-dashed line is when we take into account only the archaeological progenitors that are quiescent at the given epoch. The orange circle is for local early-type galaxies from the SPIDER survey as processed by \citet{Chan+2016}. 
{\it Right panel:} Median of the \rmass-to-\rlight\ ratios in the $r$ (orange solid line) and $g$ (cyan solid line) bands and the associated first and third percentiles for the $g$ band (light shaded area) for our CLE galaxy progenitors with $\log$(\ms($z$)/\msun)$>10.3$ \msun\ versus results for quiescent or early-type galaxies from observations at different $z'$s for the NIR-to-optical size ratios as indicated in the box; see the text for details. The red horizontal line is an approximate lower limit at $0.8\lesssim z\lesssim 2$ from several observational studies, see \citet[][]{Chan+2016}. The half-light radii in the NIR bands are expected to be close to the half-mass radii. 
    }
    \label{fig:radial-ratio-evol-cosmo}
\end{figure*}

\section{Comparison with direct observations}
\label{sec:comparisons}

In the Introduction we have cited some of the 
voluminous literature about measuring sizes and densities of galaxies at different redshifts from cosmological surveys. 
Most of these observational works concluded that the half-light radius of quiescent or early-type galaxies, of similar stellar masses, are $\sim 3-4$ times smaller at $z>2$ than local ones, that is, the progenitors of local early-type or quiescent galaxies were apparently much more compact and dense in the past. However, as also discussed in the Introduction, there are several selection effects and biases that could affect these results. In particular, the change with $z$ of the color or $\Upsilon_\star$ gradients of galaxies  imply that the measured half-light radius evolution is a biased tracer of the intrinsic half-mass radius evolution of galaxies.  
\citet[][]{Suess+2019a} analysed multiband imaging of a large sample of $\sim 7000$ galaxies from the CANDELS fields and using three methods for calculating half-mass radii, have found that the 
\rmass-to-\rlight\ ratio of most of galaxies, including quiescent ones, decreases from $z=2.5$ to $z=1.0$ \citep[see also][who use the data from \citealp{Lang+2014}]{Chan+2018}. The authors conclude that the intrinsic size evolution of galaxies is slower than previously found in light. This is because the $\Upsilon_\star$ gradients became more negative with time. 
In \citet{Suess+2019b}, the authors extended their analysis to $z\approx 0.25$ by adding $\sim 9000$ CANDELS galaxies at $z\le1.0$ and found that the \rmass-to-\rlight\ ratios decrease slower or remain constant at $z\sim 1$. 

In the left panel of Figure \ref{fig:radial-ratio-evol-cosmo} we reproduce the results reported in \citet{Suess+2019b} for quiescent galaxies with masses log(\ms/\msun)$>$10.1, black dots with error bars. The strong \rmass/\rlightr\ evolution observed at $z\gtrsim 1$ appears to flatten at $z< 1$. 
We measured the \rmass-to-\rlightr\ ratio of our CLE archaeological progenitors with masses log(\ms($z$)/\msun)$>$10.1, where the masses are calculated at each $z$ as \ms($z$)=\ms$\times$ (\mspipe($z$)/\mspipe($z_{\rm obs}$)); recall that \ms\ is the mass of the observed galaxy from the NSA catalog. We plot the running medians of \rmass/\rlight\ (orange solid line) in the same redshift bins as in \citet{Suess+2019b} and the associated quartiles (shaded area). 
Our fossil record inferences for the progenitors of local CLE galaxies show, on average, a similar trend but less evolution in the \rmass-to-\rlight\ ratios than reported in \citet{Suess+2019b} for quiescent galaxies observed at different redshfits. Below we discuss some questions to consider regarding this comparison. 

\begin{enumerate}

\item The data from \citet{Suess+2019b} correspond to galaxies defined as quiescent at the observation's redshift, while in our case, the data correspond to the archaeological progenitors of local CLE (quiescent) galaxies, 
{which at large look-back times are star-forming rather than quiescent \citep[see also \citealp{Sanchez+2019} and \citealp{Peterken+2021}]{Lacerna+2020}. How to define when a galaxy quenches its SF is a widely discussed topic in the literature, see Appendix \ref{sec:quenching}. When the specific SFR or the related birthrate parameter $b$ can be estimated, a criterion can be introduced to consider a galaxy either as star-forming or quiescent (quenched at a given time). In Appendix \ref{sec:quenching} we describe the criterion used in \citet[][]{Lacerna+2020} and its implementation to the progenitors of the CLE galaxies studied here. Thus, at each \tlb\ we exclude those progenitors that are star-forming.  }
The orange vertical-dashed line in Figure \ref{fig:radial-ratio-evol-cosmo} shows the corresponding running medians of \rmass/\rlightr\ as a function of \tlb\ of only quiescent progenitors at the given epoch.  The fraction of progenitors that are quiescent at a given \tlb\ decreases rapidly below 10\% at $z\gtrsim 1$; this is why the vertical-dashed line line ends at $z\approx1.1$. The vertical-dashed line shows that the median \rmass-to-\rlightr\ ratio of quiescent progenitors of the CLE galaxies is similar or only slightly above the median for all progenitors.


\item As will be discussed in \S\S \ref{sec:limitations}, the instrumental and observational settings of the MaNGA survey 
tend to flatten any radial difference in the stellar populations, in particular, the mass-to-luminosity gradients, and hence the \rmass-to-\rlight\ ratios.  Therefore, our  \rmass-to-\rlight\ ratios plotted in Figure \ref{fig:radial-ratio-evol-cosmo} are expected to be underestimated.

\item Last but not least, as discussed in \S\S \ref{sec:mergers} below, if the observed local E galaxies underwent some dry mergers and stellar migration, then their stellar populations might have mixed partially in the radial direction. As a consequence, their radial distribution after the mixing becomes more homogeneous than it was previously. The above implies that the fossil record method could recover lower \rmass-to-\rlight\ ratios (shallower $\Upsilon_\star$ gradients) at any epoch than those that the progenitors could have had. 
Furthermore, the minor dry mergers work in the direction of increasing the galaxy externally (see Introduction), though this effect is minimized in our analysis due to the limited FoV of the MaNGA observations, especially those from the Primary+ sample.

\end{enumerate}

Despite of the differences seen in the left panel of Figure \ref{fig:radial-ratio-evol-cosmo} between the fossil record results and the direct observations from cosmological surveys, which could be due to the considerations mentioned above, the qualitative agreement within the scatter is remarkable. Have in mind that the above results come from completely different methods and galaxy samples. A key feature of our fossil record inferences is the late upturn ($z\approx 0.3-0.4$ for massive CLE galaxies) of the \rmass-to-\rlightr\ ratio on average, which we interpret as a consequence of the global (at all radii) and continuous quenching of SF to which the CLE galaxy progenitors were subject (see \S\S \ref{sec:interpretations} below).  Unfortunately, the data from the CANDELS fields used in \citet{Suess+2019b} become scarce at $z\lesssim 0.4$ so that it is unclear whether there is or not an upturn in the \rmass-to-\rlight\ ratio. 
\citet{Chan+2016} have selected a sample of luminous ($M_r\le -20.55$ mag) red galaxies from the ``Spheroids Panchromatic Investigation in Different Environmental Regions'' (SPIDER) survey of local early-type galaxies \citep{LaBarbera+2010a}, which is based on SDSS imagery ($39,946$ galaxies) and on UKIDSS-LAS imagery ($5080$ galaxies) for the NIR bands. 
For a well selected sub-sample of $\sim 3600$ SPIDER galaxies, they derived resolved stellar mass surface density maps by using an empirical $\Upsilon_\star$–($g-r)$ color relation. 
From these maps, \citet{Chan+2016} calculated mass-weighted structural parameters such as the half-mass radius. They report that the half-mass radius of their selected early-type red galaxies ($10.2\le$log(\ms/\msun)$\le11.6$) is on average $\approx 0.87$ times the half-light $r$-band radius. We plot this value in the left panel Figure \ref{fig:radial-ratio-evol-cosmo}. It agrees very well with our determinations for MaNGA CLE galaxies, suggesting, in combination with the data from \citet[][]{Suess+2019b}, that it should have been a late upturn in the evolution of the \rmass-to-\rlightr\ ratio of CLE progenitors. 

In the literature, there are several observational studies that report the sizes of early-type or quiescent galaxies in both optical and NIR (rest-frame) bands at different redshifts. In general, the larger the wavelength, the smaller is the half-light radius of all galaxies \citep[for works based on large local samples, see e.g.,][]{LaBarbera+2010b,Vulcani+2014,Lange+2015,Kawinwanichakij+2021}. 
It is well known that the NIR bands trace relatively  well the stellar mass surface density of galaxies. Therefore, the half-light radius measured in rest-frame NIR (e.g., $K$ or $H$) bands should be close to the half-mass radius of galaxies. Under this assumption, we compiled from the literature observational determinations of the ratio of NIR to optical half-light radii for early-type/quiescent galaxies at different redshifts, and compare them with our determinations of the evolution of the \rmass-to-\rlight\ ratio in the $g$ and $r$ bands. 
Most of the observational studies mentioned above are for massive galaxies, log(\ms/\msun)$\gtrsim$10.3. Therefore, we select the progenitors of the CLE galaxies that {\it at a given} $z$ have these masses (see above for how we calculate the mass at different $z$), and plot the running medians of their \rmass-to-\rlight\ ratios in the $g$ and $r$ bands, cyan and orange solid lines, respectively, in the right panel of Figure \ref{fig:radial-ratio-evol-cosmo}. The cyan shaded band shows the associated first and third quartiles for the $g$ band.

The data for local galaxies ($z<0.1$) were taken from the following papers. 
(a) \citet[][]{Lange+2015}, who used the data from the GAMA survey to determine the half-mass radii in  bands from $g$ to $K_s$ for early- and late-type galaxies. From their fit to the size--wavelength relation, we calculate that for massive early-type galaxies $R_{\rm e,K_s}/$\reg=0.875 and $R_{\rm e,K_s}/$\rer=0.895, on average. 
(b) \citet[][]{LaBarbera+2010b} report the $R_{\rm e,K}$-to-\reg\ ratio as a function of $R_{\rm e,K}$ for the large SPIDER survey of early-type galaxies (this ratio decreases as $R_{\rm e,K}$ is larger). For massive galaxies, log(\ms/\msun)$\gtrsim 10.3$, their Figure 7 shows that, on average, the ratio is $R_{\rm e,K}/$\reg$\approx 0.76\pm0.04$. 
(c) \citet[][]{Ko+2005} determined the $K$- and $V$-band \re\ of 273 local E galaxies in different environments 
The median of $R_{\rm e,K}/R_{\rm V}$ is $\approx 0.75$, with a tendency of decrease as the environment becomes denser. 

The data for $z>0.1$ galaxies were taken from the following studies. 
(a) \citet[][]{LaBarbera+2002,LaBarbera+2003}, who determined structural properties from UV to NIR of galaxies in clusters from $z=0.29$ to $z=0.64$. For two clusters, they report that the mean of $R_{\rm H}/R_{\rm V}$ (rest-frame) is $0.62\pm 0.07$ ($z=0.31$) and $0.83\pm 0.03$ ($z=0.64$); the uncertainties correspond to  the error of the mean. 
(b) \citet[][]{Marian+2018} studied 79 massive early-type galaxies in the core of a massive cluster at $z=0.44$. The mean and error of the mean of their measured (rest-frame) $H$-to-$r$ size ratios are $R_{\rm e,H}$/\rer$=0.77\pm0.02$. 
(c) \citet{Chan+2016} studied the structure in several bands of 36 massive passive galaxies in a cluster at $z=1.39$. The reported median of the (rest-frame) $H$-to-$r$ size  ratios is $R_{\rm e,H}$/\rer$=0.83$. 
(d) From a compilation of several studies of early-type/passive galaxies at high redshifts \citep[$0.8\lesssim z\lesssim 2.5$;][]{Trujillo+2007, Cassata+2010,Damjanov+2011,Delaye+2014},  \citet{Chan+2016} conclude that the rest-frame optical radii are $\sim 20\%$ larger than the rest-frame NIR radii. Therefore, the ratio of NIR to optical sizes of these galaxies is above $\sim 0.83$. 

From the right panel of Figure \ref{fig:radial-ratio-evol-cosmo}, under the assumption that the NIR half-light radii is similar to the half-mass radii, we see that our fossil record inferences are in {rough agreement within the scatter and uncertainties, with the direct observations of galaxies at different redshifts. More observational data are necessary to confirm whether there is or not an upturn in the evolution of the ratio between the effective radii in the NIR and optical bands (or in the $\Upsilon_\star$ or color gradients). }

\section{Discussion}
\label{sec:discussion}

The results from the fossil record analysis of MaNGA ``red and dead'' Ellipticals, CLEs, presented in Section \ref{sec:results} show that their mass and light structures evolve in a significantly different way. 
Following, in subsection \ref{sec:caveats} we discuss the caveats of our approach and results. Then, in subsection \ref{sec:interpretations} we speculate about the physical interpretations and implications of our results in that regards the evolution of CLE galaxies.

\begin{figure*}
	\includegraphics[width=2\columnwidth]{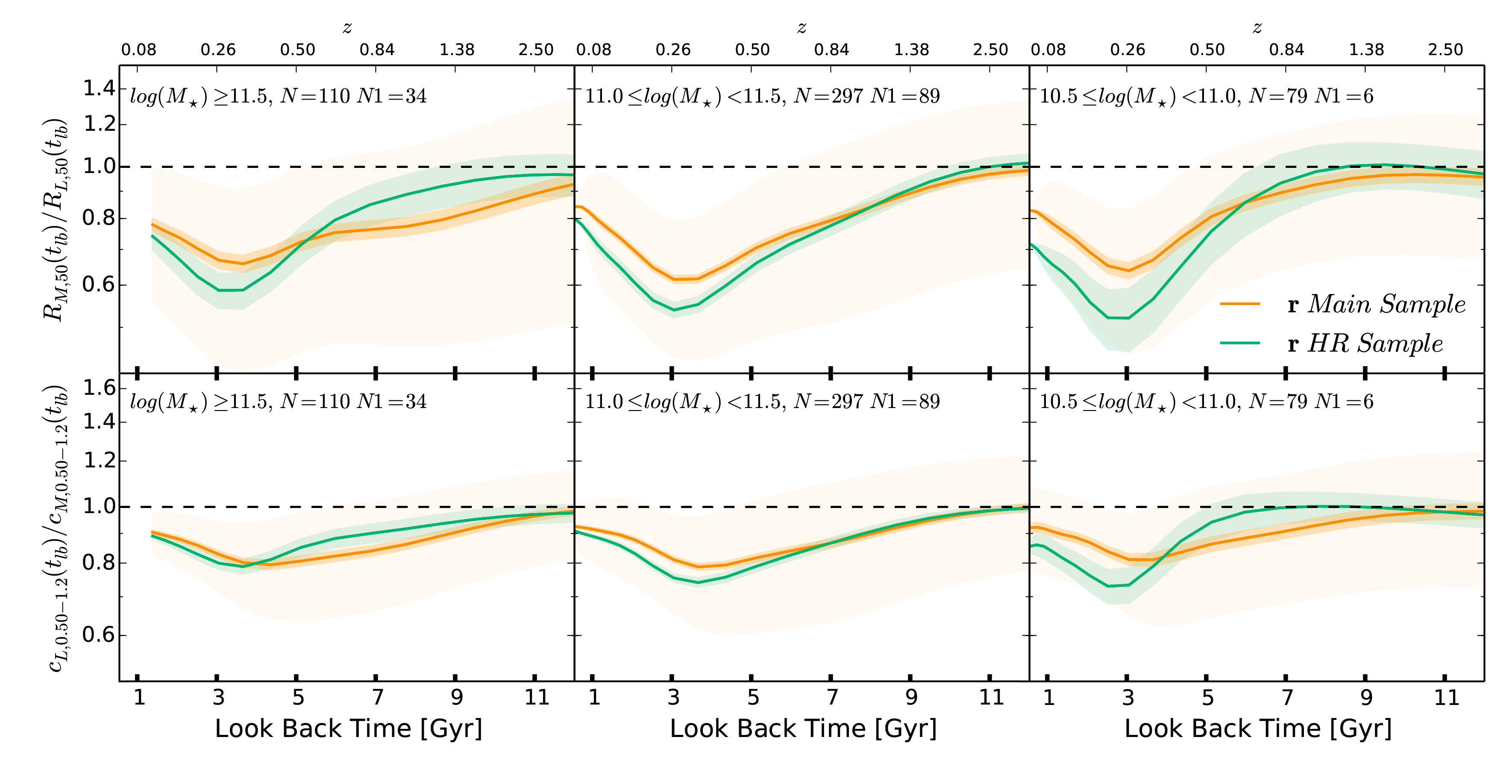}
    \caption{ {
    Comparisons of the main galaxy sample and the best spatially resolved galaxies (IFU bundles of 127 and 91 fibers) for the evolution of the ratio between the half-mass and half-light radii of the MaNGA CLE archaeological progenitors (upper panels), and the evolution of the ratio between stellar light and mass inner concentrations (lower panels). The orange color lines are for the main galaxy sample in the \textit{r} band as in Figure \ref{fig:radial-ratio-evol}. The green color lines are for the best spatially resolved galaxies in the \textit{r} band. The dark green and orange shaded regions correspond to the standard error of the median, while the light orange shaded regions correspond to the first to third quartiles (to avoid saturation in the figure, we show this only for the main sample). The \ms\ range and the number of galaxies in each bin are indicated in the top of the respective panels; N is for the main sample, and N1 for the best spatially resolved galaxies.} }
    \label{fig:radial-ratio-evol-HR}
\end{figure*}
\subsection{Caveats}
\label{sec:caveats}

\subsubsection{Methodological, instrumental, and observational limitations}
\label{sec:limitations}

The fossil record method allows us to reconstruct the age and metallicity distributions of the stellar populations in galaxies, for recent reviews see e.g., \citet[][]{Conroy2013,Wilkinson+2015,Sanchez2020}. 
This is how we calculate the stellar light and mass 2D maps corresponding to stellar populations at different look-back times for the MaNGA CLE galaxies by using the \verb|Pipe3D| code. However, the method is subject to well-known limitations and uncertainties, including those related to the spatial distributions \citep[for discussions, see e.g.,][and more references therein]{Sanchez+2016_p21,Ibarra-Medel+2016,Ibarra-Medel+2019}. Following, we discuss how these limitations could affect the results presented here. 

The precision of the fossil record method in determining the mass fractions as a function of SSP age that compose a spectrum worsen as older the true populations are due to the degeneracy in the typical spectral features corresponding to these populations. In addition to this, the SSP age binning for the stellar library used here becomes very coarse for ages 
$\gtrsim$ 8 Gyr. 
As the result, the distribution of the mass fractions is smoothed, loosing the inversion method precision in recovering the true age distribution of relatively old stellar populations \citep[][]{Ibarra-Medel+2016}. Thus, in general our results for large look-back times, \tlb$> 8-9$ Gyr, become uncertain and should be taken with caution. 


{Furthermore, the systematic biases and uncertainties of the instrumental and observational settings (number of fibers covering the galaxy, S/N ratio, galaxy inclination, etc.) introduce additional limitations in the reconstruction of the {\it radial distributions}.} \citet[][]{Ibarra-Medel+2019} applied the MaNGA settings to simulated post-processed galaxies and used Pipe3D to make global and radial archaeological inferences, as in the case  of observed galaxies. The above analysis allowed them to assess the accuracy in the recovery of the true age profiles, as well as the radial SF and stellar mass growth histories. 
The most systematical trend they have found is that radial differences in the recovered ages, SF and stellar mass growth histories become less pronounced than they actually are, that is, there is a trend to flatten out any intrinsic radial difference in the stellar populations. This trend is enhanced as lower is the spatial sampling or resolution (less number of fibers), more inclined and dust-attenuated is the observed galaxy, and lower is the S/N ratio.
However, it is important to note that the mentioned systematic trend is less significant for a bulge-dominated galaxy assembled earlier and more coherently  (as the CLE galaxies) than for a disc-dominated galaxy with a prominent inside-out growth mode.

According to the above discussion, the mass-to-light radial gradients inferred for the MaNGA CLE galaxies tend to be flattened. This implies that the \rmass-to-\rlight, $C_{0.5-1.2}^L$-to-$C_{0.5-1.2}^M$, and $\langle\Sigma\rangle_{\rm 50L}$-to-$\langle\Sigma\rangle_{\rm 50M}$ ratios presented in Section \ref{sec:results} could be underestimated, something to have in mind when comparing with direct observations of galaxies at different redshifts (see Figure \ref{fig:radial-ratio-evol-cosmo}). 
{Unfortunately, it is extremely difficult to quantify these underestimations for each galaxy as it depends on many instrumental and observational factors ''convolved'' with the particular (unknown a priori) nature of the stellar populations of a given galaxy. 
However, we can attempt to evaluate statistically how much our result can be underestimated due to the main instrumental and observational effects. According to the tests by \citet[][]{Ibarra-Medel+2019}, the accuracy in the recovery of the radial variations worsens mainly as poorer are the spatial resolution and S/N ratio. Therefore, in Figure \ref{fig:radial-ratio-evol-HR} we have calculated again the median \rmass/\rlight\ and $C_{0.5-1.2}^L/C_{0.5-1.2}^M$ tracks presented in Figure  \ref{fig:radial-ratio-evol}, but only for those galaxies best spatially resolved; for the way the  MaNGA survey was designed, they correspond to galaxies sampled with 127 and 91 fibers.
The above condition reduces the sample to $\sim 10$\%. For galaxies less massive than \ms$\approx 3\times 10^{10}$ \msun, only a few Es were observed with 127 and 91 fibers. This is why we do not present the medians and percentiles for the low-mass bins.

Figure \ref{fig:radial-ratio-evol-HR} shows the results for the best spatially-resolved galaxies in the $r$ band, and compare with those for the whole CLE sample plotted in Figure  \ref{fig:radial-ratio-evol}.
As expected, the corresponding median \rmass/\rlight\ and $C_{0.5-1.2}^L/C_{0.5-1.2}^M$ tracks attain smaller values for most of the look-back times (and specially at the minimum) than for the whole sample; thus, the gradients result less affected when the spatial resolution is higher. }
However, the differences in the medians are actually small. 
For example, for the CLEs in the $11.0\le\log$(\ms/\msun)$<11.5$ bin, the values of the median \rmass/\rlight\ ratio at the minimum is $\approx 0.62$, while for the best resolved galaxies is $\approx 0.54$. 
The above shows that limitations in the instrumental and observational settings lead to underestimating the \rmass/\rlight\ ratios, as expected, but the underestimation is small. 

In addition to the caveats mentioned above, systematical effects in the results from the inversion method are introduced by the SPS and dust attenuation modeling, as well as the choice of the IMF. As for the dust attenuation, it is not an issue for CLE galaxies since these galaxies contain only very small fractions of dust in such a way that their attenuation effect is minor.
Regarding the IMF, we assume that it is universal in space and time. There are some pieces of evidence that the IMF of early-type galaxies is not universal between them \citep[e.g.,][and more references therein]{vanDokkum-Conroy2010,LaBarbera+2013,Spiniello+2014}.
However, what is more relevant to our results is whether or not the IMF changes systematically with radius within E galaxies. Some observational works have found evidence in individual massive early-type galaxies of bottom-heavy IMFs (such as the Salpeter one or heavier) in the centre and bottom-light IMFs (such as the Chabrier one) at larger galactocentric radii \citep[][and more references therein]{MartinNavarro+2015,vanDokkum+2017,Sarzi+2018,LaBarbera+2019}, while other works do not find significant IMF radial variations \citep{Zieleniewski+2017,Alton+2018,Vaughan+2018,Feldmeier-Krause+2021}. 
More related to our study for E galaxies from the MaNGA survey, \citet[][]{Parikh+2019}, \citet[][]{DominguezSanchez+2019}, and \citet[][]{Zhou+2019} also find, in general, the trend mentioned above for the IMF with radius, though the results at a quantitative level change depending on several methodological assumptions as well as on the velocity dispersion and mass of the galaxies. The spectral inversion method for archaeological inferences is complex and relaxing the assumption of constant IMF within galaxies introduces strong uncertainties in the results. On the other hand, the use of line-index strengths, while it provides valuable clues to the IMF, may be affected by several degeneracies. For this reason, the question of IMF variation in galaxies is a highly debatable issue with sometimes contradictory results;  for a discussion, see e.g. \citet{Nipoti+2020} and more references therein.

If the observed CLE galaxies had the above mentioned gradient in the IMF, then based on the results from \citet[][see their figures 16, 17 and 18]{DominguezSanchez+2019}, we expect that the gradients in $\Upsilon_\star$ (and therefore, in our \rmass-to-\rlight\ ratios) tend to be underestimated, that is, the \rmass-to-\rlight\ ratios could be lower than what is determined here. Given that there is a certain tendency for older populations to have more bottom-heavy IMFs (close to the Salpeter IMF), we expect the \rmass-to-\rlight\ ratios archaeologically determined at high look-back times (old populations) to be less affected than at low look-back times, when the fraction of young populations increases. 
On the other hand, according to these and other authors (see above), the lower the mass of the E galaxies, the flatter the gradient in the IMF, so the main effect of a varying IMF within the galaxy is expected for the most massive CLE galaxies.



Finally, the use of other SSP stellar libraries, specially those that allow for lower metallicities than the used here (the gsd156 one, see \S\S \ref{S_ssp}), and of different SSP age samplings can lead to different archaeological results  
{at the quantitative level. However, the results are expected to remain similar at the qualitative level. For instance, in \citet{Sanchez+2016_p21,Sanchez+2016_p171}, it was shown how the \verb|Pipe3D| code recovers roughly similar properties of the stellar populations using different SSP templates, including the one adopted here.}
Moreover, the retired Ellipticals are expected to be the least affected by implementing other libraries and SSP age sampling as their stellar populations, on the one hand, have typically metallicities higher than the limit allowed by the gsd156 library, and in the other hand, they are older and more homogeneous than those of late-type galaxies. 

\subsubsection{Effects of stellar migration and mergers}
\label{sec:mergers}

A concern when attempting to reconstruct the radial archaeological evolution of galaxies is that the observed stellar populations at a given galactocentric position could have radially migrated from an earlier radial position or even formed in other galaxy that merged with the main progenitor (ex situ stellar populations). Different migration processes within galaxies were proposed, mainly for disks. 
However, there is a debate about how significant are the net radial displacements produced by these processes \citep[see][and more references therein]{Ibarra-Medel+2016}.  Several works have shown that net radial migration, mostly outwards, is small and only significant in the outer parts of galaxies, beyond two to three scale-lengths \citep[e.g.,][]{Roskar+2008, Sanchez-Blazquez+2009,DiMatteo+2014, DiMatteo+2015,Avila-Reese+2018}.
Radial displacements less than $1-3$ kpc are not actually relevant to the inferences obtained for MaNGA galaxies because these scales are not resolved. 
Regarding dry mergers  (see references in the Introduction), they produce three effects: (i) they add ex-situ mostly old/intermediate-age stars, mainly in the outer parts; (ii) driven by dynamical processes, they tend to radially mix the stellar populations; and (iii) again due to dynamical processes, they could produce a large-scale expansion of the primary galaxy. 
Therefore, the net effect of dry minor and major mergers 
goes in the direction of flattening radial differences in the stellar populations, though the final outcome depends on many pre-merger conditions and the number of mergers \citep[e.g.,][]{DiMatteo+2009}. 

Summarizing, both stellar migration and dry mergers work in the direction of flattening any gradient in the properties of the stellar populations within individual galaxies. Therefore, the inferences of stellar population properties that we attribute to what we call the archaeological progenitor of a given galaxy will tend to show flatter radial distributions than the real progenitor might have had if it had undergone stellar migration and/or mergers. However, as discussed above, the migration processes are not expected to be significant above the spatial resolution of MaNGA observations ($>1-3$ kpc), whereas dry mergers are expected to play role for massive early-type galaxies, but mostly in their outer parts. 

\subsubsection{Final remarks}

In conclusion, the inferences of the structural evolution of the MaNGA E galaxies are affected by fossil record and instrumental/observational limitations, as well as the effects of stellar migration and mergers. The main resulting effect is that the inferred gradients in the stellar population properties could be underestimated. However, this underestimate in most cases, is expected to be moderate, in such a way that any conclusion obtained with the fossil record method is correct at the qualitatively level. Even more, the inferences presented in Section \ref{sec:results} for CLE galaxies, which refer to the evolution of stellar mass-to-light ratios in structural properties, are more robust to possible radial displacements than the physical structural properties per se (e.g., the half-mass or half-light radii). This is because stellar populations of different ages but formed in the same spatial location are affected by spatial displacements roughly in the same way.

\subsection{Interpretation and implications of our results}
\label{sec:interpretations}

\begin{figure*}
	\includegraphics[width=1.4\columnwidth,height=0.3\textheight]{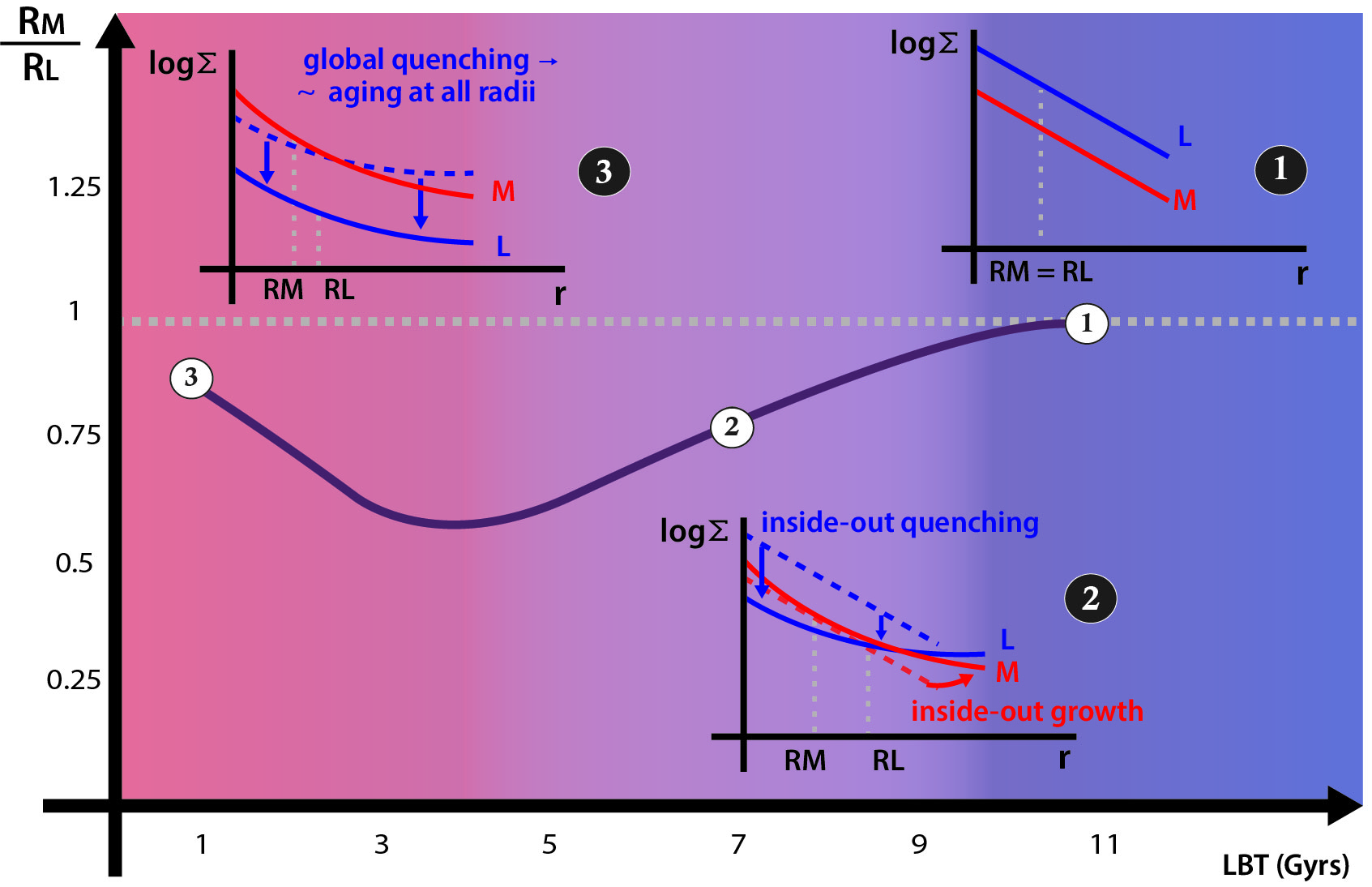} 
    \caption{{Schematic representation of the evolution of the stellar mass (M) and light (L) surface density profiles of the CLE galaxy progenitors (insets), and the resulting \rmass-to-\rlight\ ratio as a function of \tlb\ (main frame). The insets, described in detail in the text, show the current (solid lines) and previous (dashed lines) surface densities, as well as estimates of the current half-light (RL) and half-mass (RM) radii. At early epochs (bluish region), the progenitors are in the initial dissipative phase, with strong bursts of SF that with time likely concentrate more in the center due to the compaction process. At intermediate epochs (purple region), galaxies quench inside out but also grow in their external regions through dry mergers (non dissipative phase). At late epochs (reddish region), galaxies quench, age and redden even in their outermost regions (passive evolution). } }
    \label{fig:scheme}
\end{figure*}

{As mentioned in \S\S\ \ref{sec:radius-evol}, the difference in \rmass\ and \rlight\ are related to the color or $\Upsilon_\star$ radial gradients in the galaxy \citep[see e.g.,][]{Chan+2016,Suess+2019a}. In light of this, Figure \ref{fig:scheme} shows schematically how our results on the \rmass/\rlight\ archaeological evolution (Fig. \ref{fig:radial-ratio-evol}) could be interpreted (similar reasoning apply for the evolution of the $C_{0.5-1.2}^L$-to-$C_{0.5-1.2}^M$ ratio):} 

\begin{enumerate}
{ 
    \item  At early epochs, when the galaxy progenitor is in the dissipative formation phase (see Introduction), the surface brightness profile follows closely the shape of the recently built-up stellar mass surface profile in such a way that \rlight$\approx$\rmass\ (inset 1). Eventual central SF bursts due to intense gas inflow (compaction processes) can even produce temporary more compact distribution of light than mass, and then \rlight$\lesssim$\rmass.  At these early epochs, $\Upsilon_\star<1$ (Fig. \ref{fig:mass-light-evol}). 
    
    \item The SN feedback following the SF bursts and/or the activation of AGN feedback cause the exhaustion, heating or even ejection of gas, leading the galaxy to quench SF from the inside out. At the same time, the aging galaxy may grow, mainly in the outer regions, adding ex-situ stars in dry mergers with younger galaxies (non-dissipative phase).  Both inside-out quenching and the external inside-out growth lead to more extended light radial distribution than mass, i.e., \rlight\ increases with respect to \rmass\ and \rmass/\rlight$<1$ (inset 2). } 
    In fact, both inside-out quenching and inside-out growth were found for CLE galaxies in \citet[][]{Lacerna+2020}, with the same dependence on mass as reported here: the higher the \ms, the larger are the differences in the normalized mass growth and specific SFR histories between the inner ($<0.5$\re) and outer (1-1.5\re) regions, see Figs. 14 and 17 in that paper. Note that the external mass growth produces intrinsic structural evolution in the galaxy, while the inside-out quenching can happen without a significant structural change; it is more a photometric effect. In a future paper we will study which of these processes dominate in the evolution of the \rmass-to-\rlight\ (and $C_{0.5-1.2}^L$-to-$C_{0.5-1.2}^M$) ratio reported here. 
    
    \item {The upturn in the decline of the \rmass-to-\rlight\ ratio at late times has been explained in \S\S \ref{sec:radius-evol} as a consequence of the global (at all radii) and continuous decline of SF, what we call long-term quenching, see Appendix \ref{sec:quenching}. As a result, the stellar populations age even in the outer radii, and the shape of the radial surface brightness distribution approximates more and more to the radial surface mass distribution (old populations) as time passes, that is, \rlight$\rightarrow$\rmass\ (inset 3). In other words, the stellar populations, even in the outermost radii, become older than the ages at which colors or the $\Upsilon_\star$ ratio remain already roughly constant notwithstanding the age, and then, the color or $\Upsilon_\star$ gradients across the galaxy tend to flatten.
   At these late epochs the galaxies are globally red and $\Upsilon_\star>>1$, see Figure \ref{fig:mass-light-evol}.  Following  \citet[][]{Lacerna+2020}, we estimate the global quenching look-back time of each of our CLE galaxies, \tq, see Appendix \ref{sec:quenching}. Figure \ref{fig:histogram-tq-tu} plots the histogram of \tq, along with the histogram of $t_{\rm min}$, the look-back time when the galaxy reaches its minimum  \rmass-to-\rlightr\ ratio. We have smoothed the individual \rmass/\rlightr\ curves to calculate their minimums. The look-back times at which the CLEs were definitively quenched roughly coincide with the look-back times at which the minimum \rmass-to-\rlightr\ ratio has been attained, in support of our interpretation. The median and the 16th-84th percentiles of $t_{\rm min}$ are 3.6 Gyr and 2.4-4.9 Gyr, respectively, while for $t_{\rm min}$, these values are 3.4 Gyr and 2.6-4.8 Gyr, respectively.   }

\end{enumerate}

\begin{figure}
	\includegraphics[width=\columnwidth]{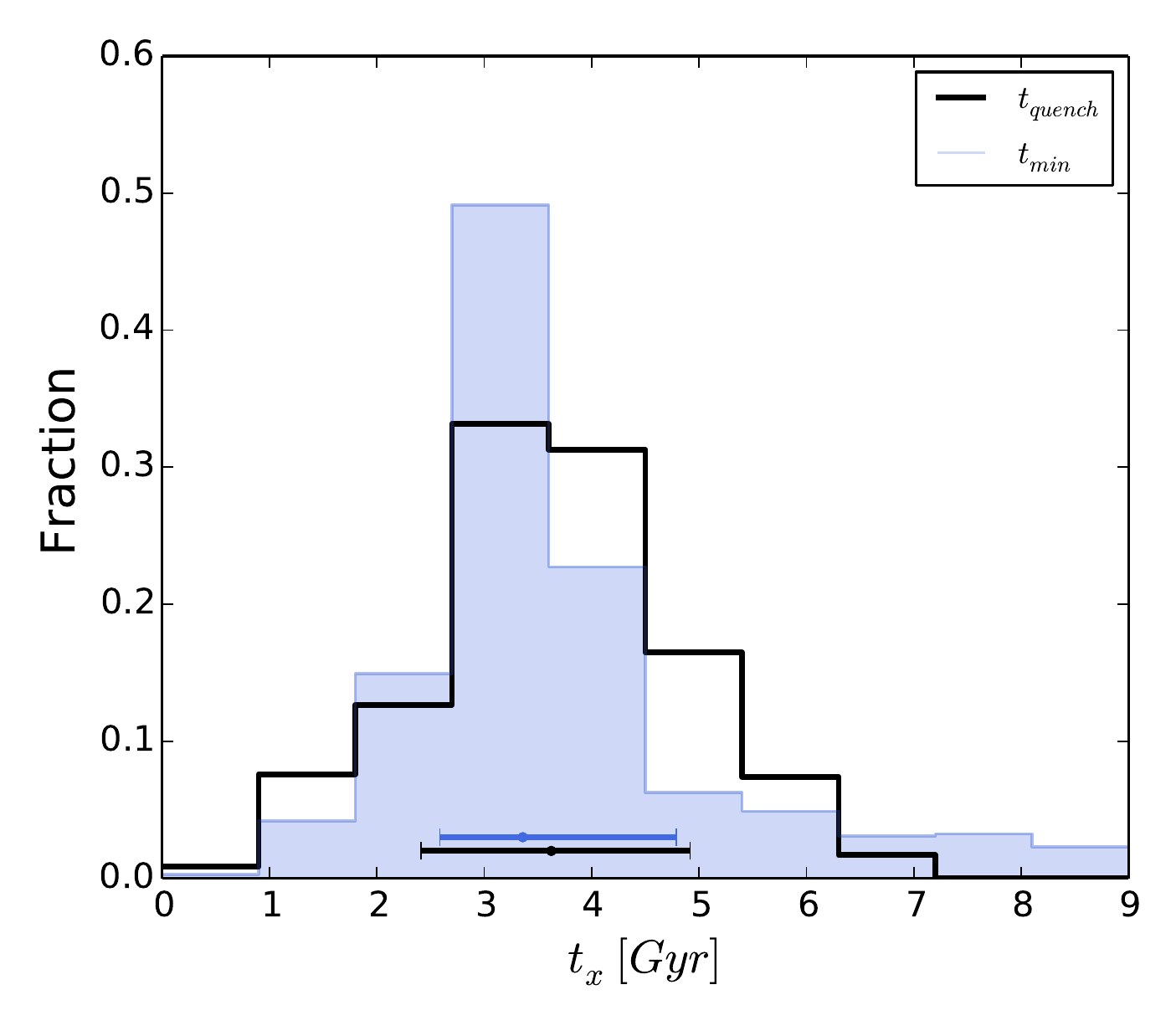}
    \caption{{Distribution of the long-term quenching look-back times of the CLE progenitors (black histogram) compared to the distribution of the look-back times when the smoothed \rmass/\rlightr\ tracks of the CLE progenitors reach a minimum (blue shaded histogram). The dots and horizontal bars show the median and 16th-84th percentiles of both look-back times, \tq\ and $t_{\rm min}$. }}
    \label{fig:histogram-tq-tu}
\end{figure}

\begin{figure}
	\includegraphics[width=\columnwidth]{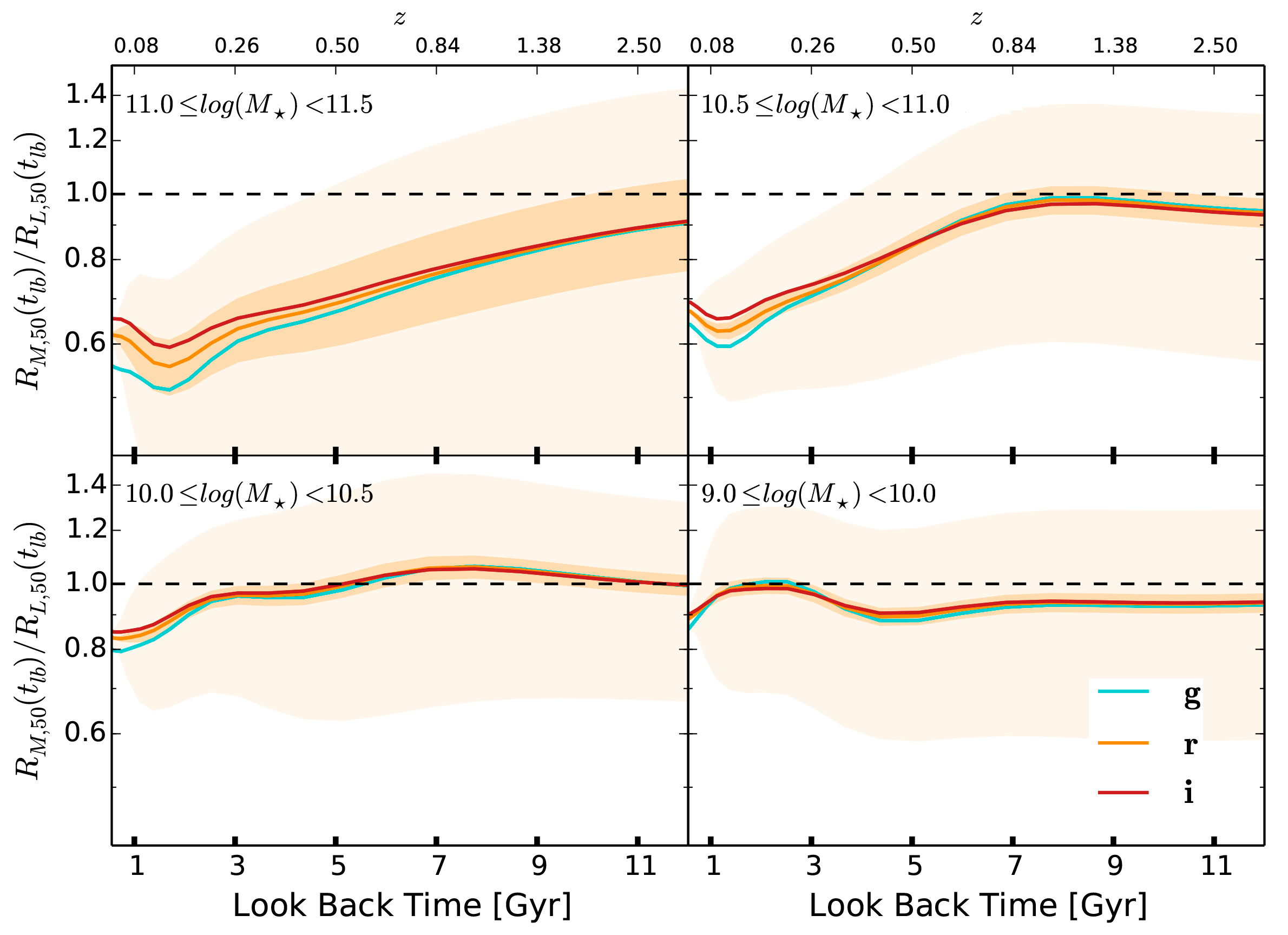}
    \caption{{Evolution of the ratio between the half-mass and half-light radii of the archaeological progenitors of MaNGA late-type star-forming galaxies ($i<60^\circ$). The line and color codes are as in Figure \ref{fig:radial-ratio-evol}. The \ms\ range in each bin are indicated in the top of the respective panels. We do not present the bin with log(\ms/\msun)$>11.5$ because there are only a few galaxies with these masses.}}
    \label{fig:LTG-SF}
\end{figure}

{To check if the upturn in the \rmass/\rlight\ archaeological evolution of the CLE galaxies is not due to some artifact of our fossil record methodology rather than the explanation given in item (iii), we also have explored the behavior of the \rmass/\rlight\ evolutionary tracks for MaNGA DR15 late-type star-forming galaxies. Galaxies later or equal than Sb according to the visual morphological classification mentioned in \S\S \ref{sec:morphology}, obeying the criteria of SF described in \S\S \ref{sec:spec-classification}, and less inclined than 60$^\circ$ were selected for the analysis. 
Figure \ref{fig:LTG-SF} shows the median archaeological evolution of \rmass/\rlight\ in four \ms\ bins, similar to Figure \ref{fig:mass-light-evol} (in the bin with log(\ms/\msun)$>11.5$ there are only a few galaxies, so we do not show this \ms\ bin). }
As expected, \rmass/\rlight\ continues decreasing up to very recent epochs (for the massive ones) or to the corresponding observation redshifts (for the less massive ones). On the one hand, these galaxies are yet star-forming (though for the more massive galaxies a significant decline in SFR may have started recently), so a strong upturn in \rmass/\rlight\ is not evident. On the other hand, the late-type disc galaxies show clear trends of continuous inside-out mass growth both from observational and theoretical studies \citep[e.g.,][]{vandenBosch1998,Avila-Reese+2000,Firmani+2009,Gonzalez-Delgado+2016,Ibarra-Medel+2016,Goddard+2017_466mass,Avila-Reese+2018,Peterken+2020} in such a way that their \rmass-to-\rlight\ ratios decrease continuously over time. The less massive late-type galaxies have in general a delayed SF history, so they can be in the initial phases of decreasing their \rmass-to-\rlight\ ratios.

{Both the \rmass/-to-\rlight\ and $C_{0.5-1.2}^L$-to-$C_{0.5-1.2}^M$ ratios characterize the differences between the stellar light and mass radial distributions.  Our results showed that their changes with \tlb\ follow similar trends (compare Figs. \ref{fig:radial-ratio-evol} and \ref{fig:concentration-evol}), though the amplitudes for the radius ratio tend to be higher than for the concentration ratio, at all \tlb\ or $z$, including $z\sim 0$. Recall that concentration is a parameter related to the inner shape of the radial distributions, while compactness is related to the average slope of these distributions \citep[][]{Andreon2020}. For local early-type or red galaxies, \citet[][]{Vulcani+2014} and \citet{Kennedy+2015}, who instead of concentration used the S\'ersic index as the shape parameter, have found that wavelength variations in effective radius are larger than variations in the S\'ersic index. Assuming that near-infrared bands trace well the stellar mass, the above results agree qualitatively with our results at low $z$. As highlighted in \citet[][]{Vulcani+2014} and \citet{Kennedy+2015}, that the wavelengths variations in \re\ are larger than variations in the inner profile's shape suggests an inner structure in Ellipticals formed by similar mechanisms that are apparent at all wavelengths, but with a global scale that increases as the band is bluer, probably due to outer growth by dry mergers. 
}

To conclude, our fossil record study of  CLE galaxies
shows that the compactness, concentration, and effective surface density of their archaeological progenitors evolve very differently between stellar mass and light. {In other words, light is a biased tracer of the intrinsic stellar structure and its evolution. An implication of the above results is that the strong structural evolution of early-type or quiescent galaxies inferred from studies based on galaxy surveys at different redshifts (see Introduction) could be, at least partially,
explained by photometric changes rather than by intrinsic structural changes
 \citep[][]{Suess+2019a,Suess+2019b}. 
 Actually, according to our archaeological analysis, the half-mass radius \rmass\ (as well as the mass concentration $C_{0.5-1.2}^M$) of CLE galaxies has changed very little with time (see Appendix \ref{appendix:R-evolution}), though these result should be taken with caution as discussed in \S\S \ref{sec:radius-evol}.
}

\section{Conclusions}
\label{sec:conclusions}

From the DR15 MaNGA survey, we have selected 537 galaxies visually classified as Ellipticals and obeying the criteria of red and dead (long-term quenched), the CLE galaxies. We have applied to this sample the fossil record method (Pipe3D code) to calculate their 2D stellar mass and light 
maps for stellar populations of different ages. Then, we have studied the differences as a function of look-back time or redshift between the stellar mass and light spatial distributions, for the $g,$ $r,$ and $i$ rest-frame bands.
To characterize these differences we have calculated the following global ratios: \rmass/\rlight\ (compactness ratio), $C_{0.5-1.2}^L$/$C_{0.5-1.2}^M$ (inner concentration ratio), and  $\langle\Sigma\rangle_{\rm 50L}/\langle\Sigma\rangle_{\rm 50M}$ (effective surface density ratio).
We have binned the results of each individual galaxy by their stellar mass at the observation redshift, \ms, and for these bins we calculated the median and scatter of the mentioned above ratios as a function of \tlb. The main results are as follows.

\begin{itemize}
    \item The \rmass/\rlight\ ratios at large \tlb\ or $z$ show a large scatter with a median value around 1. As \tlb\ decreases, this ratio decreases on average in all bands but with a trend more pronounced for the bluer ones  (Fig. \ref{fig:radial-ratio-evol}). 
    A minimum in the median \rmass/\rlight\ ratios is reached at lower redshifts, after which it increases and the scatter is reduced. These general trends differ quantitatively with mass: the larger \ms, the earlier the median \rmass/\rlight\ begins to decrease, the earlier it reaches its minimum 
    (\tlb~$\sim 4$ and $\sim 2$ Gyr in the $r$ band for galaxies in the 
    $\log$(\ms/\msun)$\ge 11.5$ and $9.0\le\log$(\ms/\msun)$<10.0$ bins, respectively), and the larger tend to be the differences between \rmass\ and \rlight. 
    
    \item The  median $C_{0.5-1.2}^L$/$C_{0.5-1.2}^M$ ratios and their scatters in the three bands show similar trends with \tlb\ and \ms\ as the \rmass-to-\rlight\ ratios, but less pronounced (Fig. \ref{fig:concentration-evol}). The latter suggests that there is less variation in the inner shapes (concentration) between the mass and luminosity radial distributions of CLE galaxy archaeological progenitors than between their respective characteristic global sizes. 
    
    \item The median $\langle\Sigma\rangle_{\rm 50L}/\langle\Sigma\rangle_{\rm 50M}$ ratios ($\approx \Upsilon_\star^{-1} (R_{\rm 50M}/R_{\rm 50L})^2$) strongly increase with \tlb\ or $z$, more for the bluer bands and more massive CLE galaxies (Fig. \ref{fig:sigmas}); {this is mainly due to the strong decrease of $\Upsilon_\star$ with \tlb\ (Fig. \ref{fig:mass-light-evol})}. The archaeological progenitors of the CLE galaxies had much higher surface densities in light than in mass in the past than in the present. 
\end{itemize}

The main conclusions of our study are that since early epochs (the more massive the galaxy, the earlier the time) the CLE galaxy archaeological progenitors \textit{(i)} become on average systematically less compact, concentrated, and dense in light than in mass as \tlb\ or $z$ are lower; however, at late epochs, \textit{(ii)} there is an upturn in these trends and the differences between mass and light in compactness or concentration decrease towards the present day, while in effective surface density, the differences continue to increase but slower than in the past.  
{We interpreted the first conclusion 
as a period of inside-out quenching and external inside-out growth after the initial phase of dissipative collapse. Both processes were found for CLE galaxies in \citet[][]{Lacerna+2020}. The inside-out quenching process can happen without significant structural changes, while the inside-out growth, driven by dry mergers, produces structural changes, mainly an external mass growth of the galaxy.  Given the limited FoV of most of MaNGA CLEs, the external growth inferred with the archaeological inference may have been only partially captured).}
Regarding the second conclusion, about the upturn in the \rmass/\rlight\ and $C_{0.5-1.2}^L$/$C_{0.5-1.2}^M$ ratios at late epochs, we interpreted this a consequence of the long-term global quenching that the CLE progenitors suffer a these epochs.
As a result, the stellar populations age even in the outermost regions in such a way that the light radial distribution tends to that one of the stellar mass distribution, which is dominated by older stellar populations. 

We have discussed that the limitations of the fossil record approach (mainly the effects of radial mixing of stellar populations by migration and mergers), as well as the MaNGA instrumental/observational settings, work mostly in the direction of flattening the radial gradients of the archaeological inferences, see \S\S \ref{sec:caveats}. 
Therefore, the trends in the different ratios between mass and light structural properties reported here are expected to be {\it underestimated}. When comparing our results on the evolution of the \rmass-to-\rlight\ ratios with these determinations but for quiescent or early-type galaxies observed directly at different redshifts, specially with \citet[][see Fig. \ref{fig:radial-ratio-evol-cosmo}]{Suess+2019a,Suess+2019b}, while the trends are qualitatively similar, our results appear to be indeed underestimates. 

Despite the methodological, observational, and interpretation limitations that our archaeological results may have (see \S\S \ref{sec:caveats}), we believe that they provide a qualitative description of the internal ($\lesssim 1.5-2$ \re) mass-to-light structural evolution of the present-day ``red and dead'' E galaxies. The above description shows that the strong structural evolution claimed for their progenitors in works based on observations of early-type or quiescent galaxies from galaxy surveys at different redshifts could be partially explained by photometric effects, {in such way that these progenitors could have evolved in a quasi-passive regime, without significant structural changes, but with a significant change over time in their mass-to-light ratios and mass-to-light radial gradients, that is, with a strong photometric evolution. In a forthcoming paper (Avila-Reese et al., in prep.) we will study the evolution of the mass-to-light and color radial gradients, as well as the inner and outer SF histories, of the CLE galaxy archaeological progenitors. }

\section*{Acknowledgements}

We would like to thank the referee for a detailed and helpful review of our manuscript. ARP and VAR acknowledge financial support from CONACyT through ``Ciencia Basica'' grant 285721, and  from DGAPA-UNAM through PAPIIT grant IA104118. We thank Casandra Rojas for her support in the preparation of Figure \ref{fig:scheme}. 

We acknowledge the SDSS-IV collaboration for making publicly available the data used in this paper. SDSS-IV is managed by the Astrophysical Research Consortium for the 
Participating Institutions of the SDSS Collaboration including the 
Brazilian Participation Group, the Carnegie Institution for Science, 
Carnegie Mellon University, the Chilean Participation Group, the French Participation Group, Harvard-Smithsonian Center for Astrophysics, 
Instituto de Astrof\'isica de Canarias, The Johns Hopkins University, 
Kavli Institute for the Physics and Mathematics of the Universe (IPMU)/University of Tokyo, 
Lawrence Berkeley National Laboratory, 
Leibniz Institut f\"ur Astrophysik Potsdam (AIP),  
Max-Planck-Institut f\"ur Astronomie (MPIA Heidelberg), 
Max-Planck-Institut f\"ur Astrophysik (MPA Garching), 
Max-Planck-Institut f\"ur Extraterrestrische Physik (MPE), 
National Astronomical Observatories of China, New Mexico State University, 
New York University, University of Notre Dame, 
Observat\'ario Nacional / MCTI, The Ohio State University, 
Pennsylvania State University, Shanghai Astronomical Observatory, 
United Kingdom Participation Group,
Universidad Nacional Aut\'onoma de M\'exico, University of Arizona, 
University of Colorado Boulder, University of Oxford, University of Portsmouth, 
University of Utah, University of Virginia, University of Washington, University of Wisconsin, 
Vanderbilt University, and Yale University.

This paper made use of the MaNGA-Pipe3D data products. We thank the IA-UNAM MaNGA team for creating this catalogue, and the CONACyT-180125 project for supporting them.

\section*{Data availability}

The data underlying this article are available at the MaNGA-Pipe3D Valued Added Catalog at https://www.sdss.org/dr15/manga/manga-data/manga-pipe3d-value-added-catalog/. The datasets were derived from sources in the public domain using the SDSS-IV MaNGA public Data Release 15, at https://www.sdss.org/dr15/,




\bibliographystyle{mnras}
\bibliography{main.bib} 




\appendix
\section{Evolution of the half-mass and half-light radii}
\label{appendix:R-evolution}

\begin{figure*}
\includegraphics[width=2\columnwidth]{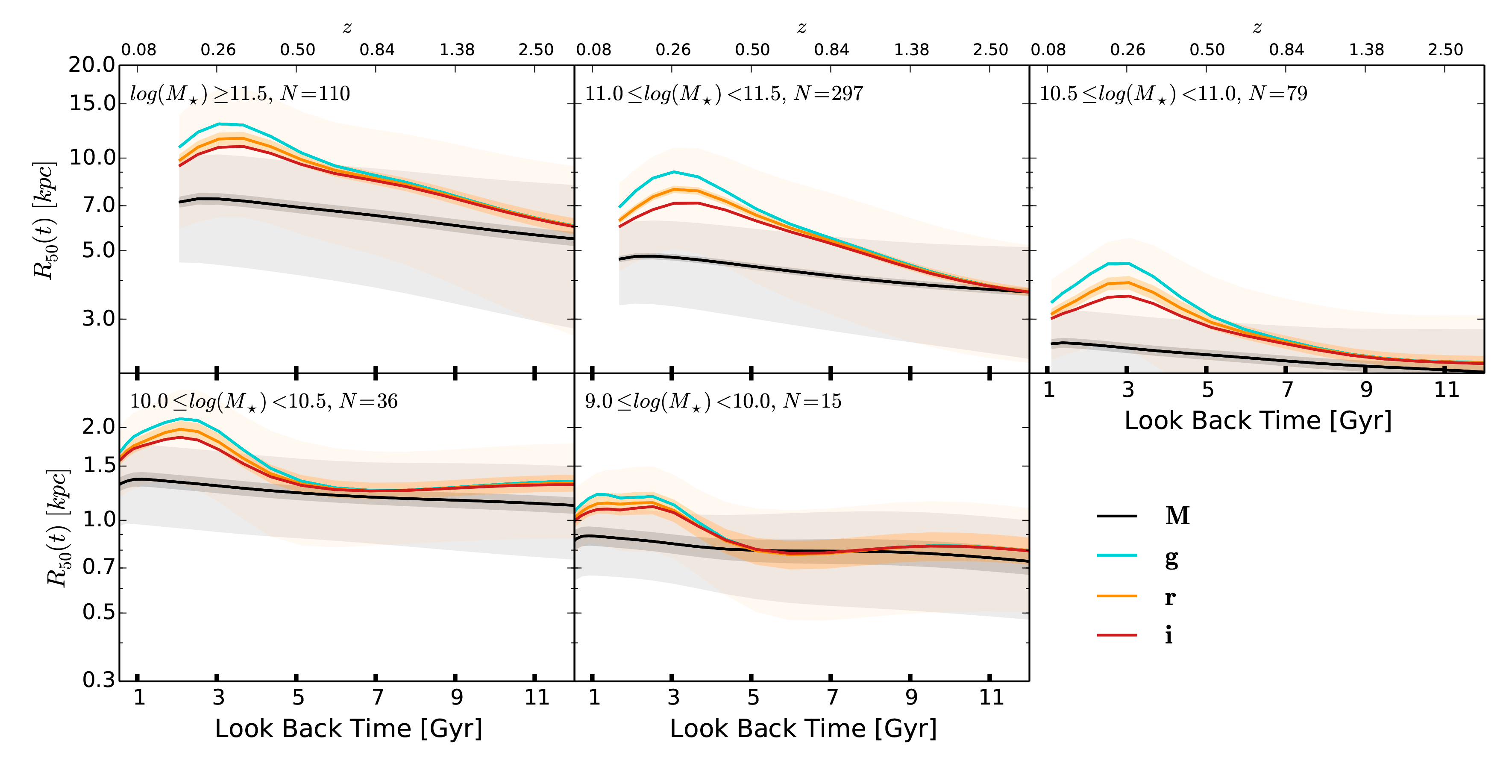}
\caption{Evolution of the half-mass and half-light radii of the MaNGA CLE archaeological progenitors. Each panel shows the running medians in different bins of \ms\ at the observation time for the $g,$ $r$, $i$ bands (cyan, orange, and red lines, respectively), and for the stellar mass (black line). The dark shaded regions are the standard error of the median, while the light shaded regions correspond to the associated first to third quartiles for the $r$ band and stellar mass (orange and black, respectively). The \ms\ range and the number of galaxies in each bin are indicated in the top of the respective panels.}
\label{fig:R-evolution}
\end{figure*} 

{ 
Figure \ref{fig:R-evolution} plots the running medians of the individual \rmass\ and \rlight{} (in the $g, r,$ and $i$ bands) of the MaNGA CLE galaxies progenitors at different \tlb, in the same \ms\ bins and with the same color code as in Figure \ref{fig:radial-ratio-evol}. The error of the mean and the quartiles for the \rmass\ and \rlightr{} tracks are shown with the dark and light shaded areas, respectively.  Note that the radii are calculated from the respective growth curves obtained non-parametrically from the Pipe3D mass and (rest-frame) luminosity maps within the FoV of each galaxy (see \S\S \ref{sec:radius-evol}). 

Half-mass radii change very little, on average, for the progenitors of the CLE galaxies. For massive galaxies (\ms$>3\times 10^{10}$ \msun), \rmass\ decrease slightly with \tlb\ (older stellar populations), but no more than by 10-20\% up to \tlb$\approx 12$ Gyr. For low-mass galaxies, \rmass\ changes even less. 
Half-light radii evolve more than half-mass radii and, on average, are always larger than the latter. 
For massive galaxies, the half-light radii in the $r$ band increase, on average, by factors $2-2.5$ from \tlb$\approx 12$ Gyr to \tlb$\approx 3-3.5$ Gyr, and then decline (there is a downturn in the \rlight\ tracks) as the stellar populations in the outer regions become older, fading in the optical bands; this makes the radial optical profiles less extended.  For less massive galaxies, the half-light radii remain roughly constant from \tlb$\approx 12$ Gyr to \tlb$\approx 5$ Gyr, then increase, and at \tlb$\approx 2-1$ Gyr they begin to decrease. Note that, given the limited FoV of most of MaNGA CLEs, their possible external growth by late dry mergers, and therefore their late size growth, could be only partially captured in our analysis. 
}

\section{A criterion for long-term SF quenching}
\label{sec:quenching} 

{
In general, by quenching we understand a process of decline of the SFR of the galaxies or their regions. However, in the literature there is not a consensus on how to define operationally SF quenching; for some recent discussions, see e.g., \cite{Schawinski+2014}, \cite{Smethurst+2015}, \cite{Pacifici+2016}, and \cite{Carnall+2018}.
We could say that a galaxy quenches as soon its SF begins to decline. However, (i) SF may gently decline by the natural aging of galaxies, and (ii) the decline may be only temporary.  As the accretion of gas decreases and its reservoir in the galaxy is consumed, SFR  will decrease. Let us call aging the cycle of decrease in gas accretion and its ulterior consumption into stars according to the  metabolism of the interstellar medium (ISM). Thus, we can understand by quenching a process of declining the SFR significantly faster than normal aging, and which is produced by some internal or external mechanisms. These mechanisms may deprive galaxies of gas infall, promote its ejection from the galaxy, or avoid the cold gas to be transformed into stars (for  reviews of different quenching mechanics, see, e.g., \citealp{Ibarra-Medel+2016}, \citealp{Smethurst+2017a}, \citealp{Hahn+2017}, and more references therein). The different quenching mechanisms have different time scales of SF cessation. On the other hand, the SF process in many galaxies may be episodic, that is, with increasing and decreasing episodes of SF. In this sense, we can say that a present-day galaxy has suffered a { long-term} quenching (it became definitively retired or passive), when its SFR no longer increases after a (final) quenching episode.

According to the photometric and spectroscopic criteria used in \S\S \ref{sec:spec-classification}, the galaxies selected as CLEs are definitively quenched (retired). At this point, we need to estimate the look-back time when they suffered its long-term quenching, \tq. Several approaches are used in the literature to determine when a galaxy is (temporary) quenched or quiescent. These approaches are based on the position of galaxies in a given color-color diagram, on the values of their specific SFR (sSFR) or birthrate parameter $b$, etc. For example, \tq\ can be defined as the cosmic time when sSFR=SFR/\ms\ decreased below a given factor $f$ times the inverse of the Hubble time \citep[e.g.,][see the latter for how this criterion connects with that based on the color-color diagram]{Firmani+2010,Pacifici+2016}, or when $b=$SFR/$\langle$SFR$\rangle$ decreased below a factor $f'$ \citep{Carnall+2018}; $\langle$SFR$\rangle=t^{-1}\int_0^t{\rm SFR}(t')dt'$ is the past average SFR.  Actually, both criteria are related, given that sSFR$(t)\approx b(t)/[(1-R_{\rm ml})t)]$, where $R_{\rm ml}$ is the fraction of stellar mass returned to the ISM; the approximation is because it is assumed that $R_{\rm ml}$ is independent of time, which is roughly correct after $\sim 2$ Gyr of evolution for a given SSP. Note that the $b$ parameter depends only on the in-situ SF history, while sSFR depends also on the ex-situ stellar mass accretion and the mass-loss factor $R_{\rm ml}$.

\citet[][]{Lacerna+2020} have estimated the long-term quenching time, \tq, for MaNGA CLE galaxies as the last of the times when $b(t)\le f'$ for each galaxy. They explored a range of $f'<1$ values, that is, when the current SFR is less than the past average SFR. They found that using $f'\approx 0.4$, on the one hand, the Blue Star-Forming Ellipticals do not attain the $b(t)$ quenching criterion at any time (if $f'\gtrsim 0.4$, then this happens for an increasing fraction of them as higher is $f'$), and on the other hand, almost all of the Recently Quenched Ellipticals attain the $b(t)$ quenching criterion at some point in the past (if $f'\lesssim 0.4$, then this does not happen for an increasing fraction of them as $f'$ is lower). For the updated sample of E galaxies used here, we find that $f'\approx 0.5$ does a better job. This is the value used to calculate the distribution of \tq\ plotted in Figure \ref{fig:histogram-tq-tu} as well as the fraction of quiescent galaxies at different look-back times in \S\S \ref{sec:comparisons}. 

}


\bsp	
\label{lastpage}
\end{document}